\documentclass[manuscript,screen]{acmart}

\AtBeginDocument{%
  \providecommand\BibTeX{{%
    \normalfont B\kern-0.5em{\scshape i\kern-0.25em b}\kern-0.8em\TeX}}}

\setcopyright{acmcopyright}
\copyrightyear{2018}
\acmYear{2018}
\acmDOI{XXXXXXX.XXXXXXX}

\acmConference[Conference 'XX]{Make sure to enter the correct
  conference title from your rights confirmation emai}{June 03,
  2018}{NY}
\acmPrice{15.00}
\acmISBN{978-1-4503-XXXX-X/18/06}

\usepackage{amsmath} 
\usepackage{newtxmath}
\usepackage{csquotes}
\usepackage{centernot}
\usepackage{caption}
\usepackage{subcaption}
\usepackage{graphicx}
\begin{document}

\title{Causal Graph for CPS Reconnaissance using Anonymized Data: A Case Study}
\title{Identify Impact of Cyber Attacks using Causal Inference on Causal Graphs of Cyber Physical Systems: A Case Study}
\title{ICCPS: Impact discovery using Causal inference of cyber attacks in CPSs}

\author{Rajib Ranjan Maiti}
\email{rajibrm@hyderabad.bits-pilani.ac.in}
\orcid{0000-0002-5510-8217}
\affiliation{%
  \institution{Birla Institute of Technology and Science, Pilani, Hyderabad Campus}
  \streetaddress{Hyderabad Campus}
  \city{Jawahar Nagar}
  \state{Hyderabad}
  \country{India}
  \postcode{500078}
}
\author{Sridhar Adepu}
\email{sridhar.adepu@bristol.ac.uk}
\affiliation{%
  \institution{University of Bristol}
  \state{Bristol}
  \country{UK}
  \postcode{}
}

\author{Emil C. Lupu}
\email{e.c.lupu@imperial.ac.uk}
\affiliation{%
  \institution{Imperial College London}
  \country{UK}
  \postcode{}
}

\begin{abstract}
  We aim to address the problem of quantifying the impact of a cyber attack in Cyber Physical Systems (CPSs).  
  In particular, we provide a solution to identify the exact set of Design Parameter (DPs) that are affected due to a cyber attack launched on another set of DPs in a CPS.
  We adopt the concepts of causal graphs into the domain of CPS to causally link one DP with another and \emph{quantify} the causal impact of one on another. 
  Using SWaT, a real world testbed of CPS on water treatment system, we demonstrate that causal graphs can be build in two ways, using i) domain knowledge of the control logic and the physical connectivity structure of the DPs in a CPS, we call them as \emph{causal domain graphs} and ii) operational data logs, we call them as \emph{causal learnt graphs}. 
  Then, we have compared these two types of graphs in terms of the edges when a same set of DPs is used in both the cases. 
  Our analysis shows that the edges need not be mutually exclusive in the causal domain graphs and the causal learnt graphs; in fact, there is a significant overlap of the edges.
  We have used the causal domain graphs for an estimation of the parameters of the graphs, whereas the causal learnt graphs for causal inference. 
  We have used three causal graph structure learning algorithms, Peter Clarke (PC), Hill Climb (HC) search and Chow-Lie (CH), to learn the causal graphs in all six-stages of SWaT. 
  Finally, we demonstrate an application and use of the causal graphs with respect to a set of well known nine cyber attacks, where the DPs impacted due to each of the attacks are discovered with a probability higher than 0.9 using causal inferences on the causal learnt graphs. 
\end{abstract}

\ccsdesc[500]{Computer systems organization~Embedded and cyber-physical systems}
\ccsdesc{Security and privacy~Systems security}
\ccsdesc[100]{Networks~Network reliability}

\keywords{Cyber Physical Systems Security, Industrial Control Systems Security, Causal Graphs, Causal Inference in CPS, Quantify the Impact of Cyber Attack}

 

\maketitle

\section{Introduction}
\label{sec:intro} 
Cyber security in Cyber Physical Systems (CPSs) has attracted a significant amount of research in recent past and such popularity is primarily driven by the impact of reported cyber incidents \cite{miller2021looking} on such systems. 
Certain studies, like \cite{adepu2019challenges}, have conjectured that any effective defence mechanism against cyber attacks in CPSs should combine both safety (managing harm due to successful attacks) and security (protecting data and services from cyber attacks) measures. 
In this paper, we are interested to address a problem of quantifying the impact of cyber attacks on a CPS. 
Our solution involves a novel approach of causal graph, where we distinguish the design parameters (in brief, sensors or actuators in a CPS, in short DPs) that are used to launch a cyber attack (called as Targeted DPs) from those that are impacted (called as Impacted DPs) given the attack is launched on Targeted DPs.
Essentially, we believe that if an attacker exploits a vulnerability in a particular DP, say $DP_i$, (i.e., a Targeted DP), then it impact is can be observed on another DP, say $DP_j$, (i.e., an Impacted DP) due the structural coupling and control dependency inherent in a CPS. 
Therefore, we address the problem of discovering and quantifying all the potential DPs that belong to the Impacted DPs given that a cyber attack is launched on the Targeted DPs. 




In general, a CPS is integrated with a \emph{historian} that collects time series data of associated sensors and actuators that are part of the CPS. 
Such a data log, we call as historian data log, contains the value sensed by each sensor and the state of each actuator at each time step, where it is assumed that all the sensors and actuators are time synchronized. 
Because historian data log is expected to be used for additional analytics, like state space estimation \cite{ProkopevCPSADM2020, RamamurthyICCA2014, DingTSMCS2021}, anomaly detection \cite{YuanACMSurvey2021}, a CPS authority can share these data logs with a third party, in addition to in-house analysis, for an inspection of security incidents. 
One such cyber security analytics can be facilitated by an investigation of \emph{causal} dependency among a set of DPs from the time series data of historian. 
We consider each hardware component that is used for either sensing or actuating, e.g., a pH sensor or a motorised valve, a design parameter(DP) in a CPS. 
Hence, if a CPS contains 5 pH sensors, 7 motorised valves and 9 pumps, then we consider the CPS to have a total of 5+7+9 = 21 DPs. 
It is not necessary that a published data set from a historian contains all the DPs in a CPS, i.e., an owner may decide to publish a subset of the data logs.  
Note that the data log need not contain the exact mapping between the identifiers of a DP, like AIT101, and the functional description of the DP, like pH sensor.
Informally, causal dependency is a directional relationship from one DP $DP_i$ to another DP $DP_j$ in a CPS, where a change (not necessarily a linear change) in the vales of $DP_i$ causes a change in the values of $DP_j$. 
We expect that the causal dependency between two DPs can efficiently reveal the impact of a cyber attack on a DP on another DP if the DPs are causally dependent with a high probability.

\textbf{Motivation: }The work in this paper is motivated by a fact that an attack can be launched by exploiting a vulnerability of a particular DP and its impact can be observed on a different set of DPs in a CPS. 
Further, discovering the DPs that are impacted \emph{due to} an attack and quantifying the impact can lead to strengthen the security mechanisms in a a CPS. 
For instance,  if a pH sensor is attacked in such a way that it indicates a value of 6.5, instead of its real value 7.5, then the pumps that inject sodium carbonate and sodium hydroxide can be switched on so that the pH level in water can be increased. 
Therefore, it is necessary to identify the DPs that can be affected \emph{due to} an attack that is launched on a particular DP(s) representing such pumps. 
Note that due to privacy, safety and security reasons, exact mapping between the identifiers of DPs and the functionalities of sensors or actuators may not be revealed along with the historian data logs.
For instance, an identifier AIT202 indicates a DP, but its functionality is not present in the historian data log; however a plant engineer can have the knowledge of its functionality.
Hence, we aim to discover the DPs and associate a level of confidence to each of those DPs that are affected due to an attack on a different set of DPs in a CPS.


\textbf{Our Work: }We have considered a real testbed, called SWaT, of CPS that is the scaled down version of a water treatment plant having a capacity of producing five gallons of drinking water per hour. 
Historian of SWaT collects data at an interval of one second. 
We have consider a sample of about \textbf{1 GB data} of SWaT that has been released for research.  
The data log contains 51 DPs that provide only symbolic identifiers. 
The exact functionality of any of these DPs is not revealed. 
We have used this data to perform two primary task related to causal graphs. 
In brief, a causal graph is a directed acyclic graph, where each node $v_i$ indicates a random variable corresponding to a DP in the CPS and an edge exists from a node $v_i$ to another node $v_j$, if the DP corresponding to $v_j$ is causally dependent on the DP corresponding to $v_i$.
A causal relationship from $v_i$ to $v_j$ can be derived in two ways. 
One, by analysing the functional relationships among the DPs when a CPS is being developed and not yet operational. 
Two, by analysing the dynamics in the observed values of the DPs when a CPS is operational.  
Thus, we consider building causal graphs in both of these ways and use the data set to achieve two different goals. 

First, we have constructed a set of causal graphs  using the domain knowledge, where we have exploited the relationships of \emph{control dependency} and \emph{physical coupling} to form the edges; let us call these graphs as \emph{causal domain graphs}.
The data set is then used to learn parameters of these domain graphs in order to understand the properties of the graph. 
Second, the data set is directly used to learn the causal relationships between the DPs, which leads to the learning of causal graphs in a CPS; let us call these graphs as \emph{causal learnt graphs}.  
Finally, we have developed a simple algorithm that uses causal inference to identify the impact of a cyber attack when a learnt causal graph is taken as input. 
Notably, the algorithm makes use of the parameters learned in case of the causal domain graphs. 
We have used a set of nine well known cyber attacks on SWaT and we have identified their impact to demonstrate the application and use of causal graphs. 

We summarise our contributions as:
\begin{itemize}
    \item We have adopted the concepts of causal dependency to formulate causal relationships among the DPs in order to build causal domain graphs for given set of DPs in a CPS being developed. 
    We have shown how to convert the functional relationships like control dependency and physical coupling to causal dependencies in order to build the causal domain graphs. 
    Our analysis of the estimated values of the parameters of such a graph shows that the probability of each of the edges ($v_i,v_j$) in the graph is more than 0.95 in general. 
    
    \item We show how to convert each of the continuous random variables in a historian data log to a suitable discrete random variable in order to employ causal structure discovery algorithms in the causal learnt graphs. 
    Our analysis shows that there is a non-negligible number of common edges between the causal domain graphs and the causal learnt graphs using both Peter Clark (PC) and Hill Climb (HC) search algorithms.
    In addition, each of these learning algorithms can discover non-trivial and not-seen-before causal relationships when compared with the corresponding causal domain graphs.
    
    \item We have used nine well known cyber attacks on SWAT to evaluate our causal graphs. We show that, in 7 out of the 9 attacks, the impacted DPs due to an attack on the targeted DPs can be determined with a probability higher than 0.90. 
    Also, in 4 out of 9 attacks, domain knowledge cannot help to discover the impacted DPs, whereas the causal learnt graphs using PC and HC algorithms can determine their impacts.
\end{itemize}













\section{Background}
\label{sec:background}
In this section, we briefly describe the concepts of statistically (in)dependent events, statistical causality,  basics in causal graphs, and finally, the CPS, SWaT, that is used as a case study in this paper. 

\subsection{Conditional Probability and Causality}
Two related concepts of correlation and dependence are well studied in a wide range of disciplines including Physics, Philosophy, Experimental Design and Statistics \cite{pabloSpringer2012, HuiningANIPS2014, EliasPNAS2016, eliASI2018, RuochengACMSurvey2021}. 
The concepts can be better understood or distinguished by using simple real life examples. 
Raining makes roads wet -- an example of causation -- indicates that one event causes another event to hold true. 
Vehicle speed helps to reduce commute time in transportation -- an example of correlation -- indicates that both the events are related to each other. 
Correlation is usually a linear relationship between two variables, i.e., $y=mx +c$, where $m$ and $c$ are parameters of the linear relationship \cite{corrandcoef}. 
Essentially, a relationship that is not linear cannot be captured using correlation, e.g., $y=ax^2+bx+c$, where $a,b,c$ are the parameters and $x\in \mathbb{R}$ \cite{polyregress}. 

Statistically, there is a number of measures in each of correlation and causation.
For instance, Pearson correlation indicates if two random variables are positively (tending to 1) or negatively correlated (tending to -1), or no correlated (tending to 0). 
Causation can be \emph{estimated}, as opposed to measured in case of correlation, using conditional probability and conditional independence \cite{pittphilsci16732}. 
Note that causation can be unidirectional or bidirectional and bidirectional causation need not imply a correlation.  
Also, because two random variables show a high (positive/negative) correlation, the variables need not have a (unidirectional/bidirectional) causal relation.  
Statistically, the presence of a causal relation is indicated by the negation of conditional independence between two random variables \cite{ChristopherStandfordEncyclo2021}. 

Consider two discrete random variables $X_i$ and $X_j$, having $n_i$ and $n_j$ discrete values respectively.   
$Pr(X_j = x_l|X_i = x_k)$ indicates the conditional probability that $X_j$ takes the value $x_l$ given that $X_i$ has the value $x_k$. 
For any pair of values of $X_i$ and $X_j$, if conditional independence does not hold, i.e., $Pr(x_l|x_k) \neq Pr(x_l)$, then we can say that the random variables are dependent. 
If the conditional dependence holds for $Pr(X_j|X_i)$ then $X_j$ is dependent on $X_i$. Alternately, a change in the value of $X_i$ causes a change in the value of $X_j$ and not the other way around. 

To comprehend causal relationship, we need a causal model that describes causal relations among variables using a mathematical abstraction. 
Causal model can represent already known causal relations  and it can be used to discover or learn missing relations from the corresponding data.
One such causal model is Structural Causal Model (SCM) that consists of two components: i) causal graph and ii) structural equation \cite{TomerArxiv2020, MatthewACMSurvey2022}.
A causal graph is a directed acyclic graph whose vertex set represents the set of random variables and the edge set the causal relations between each pair of random variables. 
A directed edge in the causal graph from a vertex $v_i$ to a vertex $v_j$ indicates that a change in $v_i$ causes a modification in $v_j$; $v_i$ is called parent of $v_j$ in the directed graph. 
Statistically, $Pr(x_j|x_i)$ is represented by the directed edge $(x_i,x_j)$ in the graph.   
A parent vertex can have multiple children and a child can have multiple parents in the graph. However, no direct cycle can be present in the graph to qualify it as a causal graph. 
A structural equation helps to quantify the causal effect of one or more variables on another variable. 
These equations can be used to estimate the causal affects using appropriate data set for the random variables represented in the graph. 

The studies on causal graphs can be divided into two parts: 
\begin{enumerate}
    \item Causal Discovery - answers the question of which set of variables need to be modified in order to change the values of a specific variable.   
    \item Causal Inference - answers the question of how much change need to be introduced into a particular variable(s) so that an expected level of manipulation can be observed in the value of another specific variable.  
\end{enumerate}
While causal discovery helps to build a causal graph from data samples, causal inference allows the analysis of causal graph to discover causal relations that were previously not known \cite{SpirtesAppliedInfo2016}.  

\subsection{Basic Concepts in Causal Graphs}
Assume that we have a causal graph as shown in Figure \ref{fig:causal-graph-concepts} containing eight nodes, i.e., $V(G) = \{x_1, x_2, ..., x_8)$ and eight edges. 
We highlight four sub-graphs, indicated by green oval, blue rectangle, black triangle and pink diamond, that are used to define different concepts in the domain of causal graph. 
Let us consider a few of them:
\begin{figure}[!t]
  \centering
  \includegraphics[width=0.6\linewidth ]{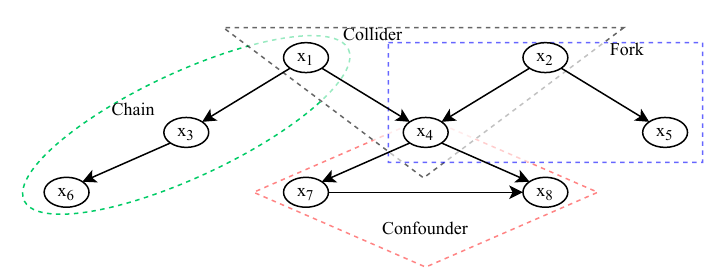}
  \caption{A simple causal graph with 8 nodes.}
  \label{fig:causal-graph-concepts}
\end{figure}

\begin{itemize}
    \item \textbf{Path:} A sequence of adjacent nodes, without repeating any edge and without regard of the directions of the edges. While an edge represents a direct causal influence, a path can represent an indirect causal influence. 
    
    \item \textbf{Directed Path:} A simple path (i.e., a path without repetition of any vertex) from a source node $v_i$ to a destination node $v_j$, where the directions of the edges allow a traversing from $v_i$ to $v_j$. 
    For instance, $\{x_1, x_3, x_6\}$ denotes a directed path from $x_1$ to $x_6$ in the graph in Figure \ref{fig:causal-graph-concepts}. 
    A directed path indicates the propagation of causal influences along the nodes in the path. 
    \item \textbf{Chain:} A particular directed path that used for propagating causal effects from a source node, to a destination node. 
    For instance, the sub-graph indicated by green oval is an example of a chain that can be used for analysing causal affect of $x_1$ on $x_6$.
    
    \item \textbf{Fork:} A node that have two or more children, indicating a common cause to two or more variables. 
    For instance, the sub-graph indicated by blue rectangle is a fork where $x_2$ is a common cause for both $x_4$ and $x_5$. 
    
    \item \textbf{Collider:} A node that has two or more parents, indicates that two or more variables causes a single variable. 
    For instance, the sub-graph indicated by black triangle shows a collider node $x_4$ that is causally effected by both $x_1$ and $x_2$. 
    While chain and fork propagates the flow of dependency, collider is likely to block the flow.
    
    \item \textbf{Blocked Node:} A node $v_i$ is said to be blocked by conditioning on a set $V'\subset V$ of nodes if any of two conditions hold: i) $v_i\in V'$ and $v_i$ is not a collider node, ii) $v_i\in V'$ and $v_i$ is a collider node and no descendant of $v_i$ belongs to $V'$. 
    For instance, $x_3$ in the \emph{Chain} is blocked, and $x_2$ in the \emph{Fork} are two blocked nodes. 
    The notion of blocked node is used to discover dependency separation (i.e., d-separation) in a causal graph. 
    \item \textbf{Confounder:} A variable that can predict or affect both a treatment (a variable that is altered) and an outcome (a variable that represents outcome of the alteration). 
    For instance, the sub-graph indicated by red diamond shows a confounder $x_4$ where the treatment is $x_7$ and the outcome is $x_8$. 
    
    \item \textbf{D-Separation:} Two nodes $v_i$ and $v_j$ are said to be d-separated (short form of dependency or directed separated) by a set $S$ of nodes if for every directed path $P_{ij}$ from $v_i$ to $v_j$, at last one vertex in $P_{ij}$ belongs to $S$ and $S$ blocks every such path from $v_i$ to $v_j$. 
    Informally, d-separation is used to decide whether a set $S_1$ of nodes are independent of a set $S_2$ of nodes, given another set $S_3$ of nodes in a causal graph, where $S_1$, $S_2$ and $S_3$ are mutually exclusive. 
    
    \item \textbf{Faithfulness Test:} Informally, a statistical test that helps to assess a causal graph generated out of a dataset if every conditional independence exhibited by the dataset is represented in the causal graph. 
    
    \item \textbf{Equivalence Class} - A mechanism to decide if two causal graphs encode or imply a same set of conditional independence among the variables. For instance, two subgraphs indicated by green oval and blue rectangle represent a same set of conditional independence. 
    
\end{itemize}

\subsection{SWaT: An Operational Testbed of a Cyber-Physical System}

\begin{figure*}[!t]
  \centering
  \includegraphics[width=0.8\linewidth]{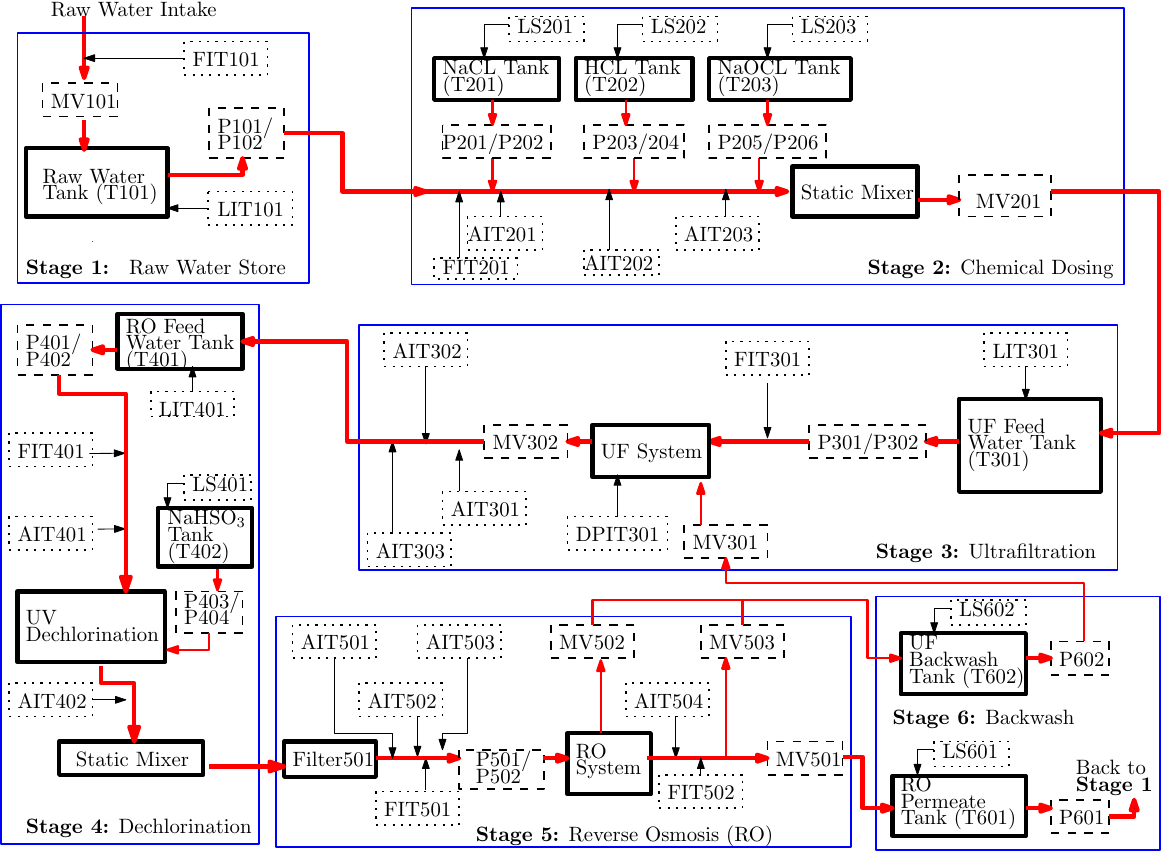}
  \caption{Six-stage cyber-physical design of secure water treatment plant. The thick rectangles are the traditional physical components of the water treatment plant and all other rectangles are the parts of cyber-infrastructure of it. Thick red arrows indicate water flow directions in the plant. Dotted- and dashed-rectangles indicate sensors and actuators respectively.}
  \label{fig:swat_processes}
\end{figure*}


SWaT (Secure Water Treatment Plant)\,\cite{mathur2016swat} is a scaled down version of a real water treatment plant having capacity of producing 5 gallons of drinking water per hour. 
Figure \ref{fig:swat_processes} shows a schematic diagram of SWaT containing important sensing and actuating hardware components in it. 
The complete physical process of SWaT is divided into six stages, each consisting a sub-process in a real water treatment plant. 
i) Stage-1: Raw water store, ii) Stage-2: Chemical dosing, iii) Stage-3: Ultrafiltration, iv) Stage-4: Dechlorinatin, v) Stage-5: Reverse Osmosis and vi) Stage-6: Backwash.  
We are interested in the hardware components that are used as either sensors or actuators in each of these stages and we call each of these components as design parameter (DP), i.e., a DP is either a sensor or an actuator. 
Each DP is identified by an alpha-numeric string where the prefix alphabetic string is used to indicate a type of functionality and the numeric string is used to indicate the stage number in which it is used along with a sequence number in that type of DPs.
For instance, "MV201" is a DP that is a motorised valve used in Stage-2 and it is first such valve in this stage. 
Similarly, "MV302" is a motorised valve used in Stage-3 and it is second valve in this stage.
For the purpose of the work in this paper, we are not interested in the sequence number, like first or second, of a DP.
Rather, we shall be focusing on the how the DPs can be associated with one another within a stage or across stages.

A sensor produces real values and an actuator integer values. 
In general, sensors are used to measure the quantity of water or the quality of water in terms of the amount of certain types of chemicals, like pH and chlorine. 
Any DP that is prefixed by AIT is used for sensing chemical properties, FIT for sensing water flow rate, LIT for sensing water level in tanks, DPIT for sensing difference in water pressure and PIT for sensing water pressure in a pipe. 

Actuators are used to control flow direction or pumping of water. 
Any DP that is prefixed by MV is used for controlling the direction of water flow, P is used for water pump and LS is used for level indicator of chemical in chemical tank. 
For a simplistic view of SWaT, we have not represented all the DPs that are present in the plant. 
However, we have maintained a consistent physical location of the DPs in each of the stages to indicate a kind of physical coupling present among the DPs in SWaT. 
For instance, in Stage-1, MV101 and FIT101 are placed before the Raw Water Tank T101 to indicate that both of these DPs are actually located on the main water pipe (indicated by red thick line) that is used to push water into T101.

\subsection{Event and Random Variable in SWaT}
In general, an event in a software indicates a change in the value of a critical variable and the change triggers an action. 
Extending the concept, an event in CPS is defined in both the physical and soft domain \cite{WangWorkshop2014}. 
In the physical domain, an event can be defined easily in case of an actuator, because it can assume a small number of values/states; whereas, in case of a sensor, defining an event can be relatively complex, because it assumes a value within a range of real numbers. 
We consider that a range of sensor values can be divided into a set of intervals and an event is change in the intervals. 

In SWaT, an event in case of MV101 being an actuator is that it changes its state from Close to Open.
Similarly, in case of LIT101 being a sensor, an event is that it changes its value from an interval [100, 210] to an interval [210, 750]. 
For simplicity, we associate a name to each such  interval, like Low, Medium or High, based on the operating principle of SWaT. 
Note that LIT101 is a sensor used to indicate water level in a water tank and its value range if 100\,L to 1000\,L. 

\section{Proposed System}
\label{sec:sys-model}
This section describes the system model used in this paper for building and applying causal graphs.
We consider that a CPS can use its historian data log for advanced services like identifying Targeted DPs and discovering Impacted DPs of cyber attack. 
Specifically, we propose to apply a causal graph to quantify the impact of a cyber attack on a CPS.

\subsection{System Model}

\begin{figure}[hbt!]
  \centering
  \includegraphics[width=0.8\textwidth]{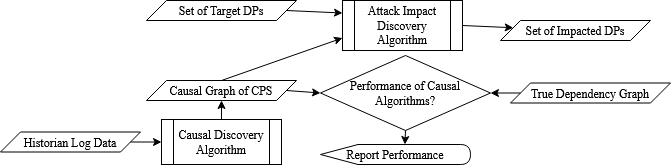}
  \caption{The architecture of our proposed system called ICCPS.}
  \label{fig:sys-model}
\end{figure}

Our system model considers two inputs: first, a large sample of historian data log and second, a set of DPs on which a certain cyber attack is launched, called Target DPs and denoted by $\mathcal{A}_{s}$. 
Historian data logs are used by the causal discovery algorithm as input, and the target DPs are used by impact discovery algorithm as input. 
The impact discovery algorithm takes causal graph as another input, that is generated by the causal discovery algorithm. 
Note that the causal discovery algorithms are used to learn the causal structure among a set of DPS in a CPS.
We add an optional step of contrasting the causal learnt graphs with the causal domain graphs. 
Building the causal domain graph is a non-trivial task, because the relationships among the DPs are not explicitly meant to be causal relations in a CPS. 
Rather, the relationships among the DPs in a CPS are primarily control dependency or physical coupling due to their installation points in the CPS. 

Let us denote by $\mathcal{A}_s$ the DPs on which a cyber attack is launched, i.e., $\mathcal{A}_s$ is a set of Targeted DPs. 
Let us denote by $\mathcal{A}_{d|s}$ the DPs that are are impacted due to an attack on $\mathcal{A}_s$, i.e., $\mathcal{A}_{d|s}$ is a set of Impacted DPs. 
Note that $\mathcal{A}_{s}$ is by default a part of $A_{d|s}$ and hence we shall consider only those DPs that are explicitly present in $A_{d|s}$ for a given cyber attack. 
We propose to quantify each DP in $A_{d|s}$ with a probability associated with it indicating a level of confidence that the DP is impacted. 
If the probability is greater than a threshold then the DP is considered as impacted with a high confidence due to the cyber attack. 
The threshold probability is decided empirically based on the parameters of the causal domain graphs. 
Thus, the output of our proposed system is set of DPs with their associated probabilities as Impacted DPs of a cyber attack that is launched on Targeted DPs. 
In summary, the aim of our system is to aid in the identification of  the DPs affected by an attack based on an analysis of the operational data logs.  

\subsection{Attacker Model}

We assume that an attacker is \emph{curious} to inspect the impact of a certain set of cyber attacks and has the ability to compromise an arbitrary set of DPs in a CPS. 
The attacker can choose any number of DPs as its target, however the objective is to look for a small set of DPs as Targeted DPs and see if Impacted DPs can be significantly different from the Targeted DPs. 
Alternately, the attacker can wish to harm with a high confidence a set of DPs without considering them as part of Targeted DPs. 
The attacker has the ability to synchronize DPs in the set of Targeted DPs, if required, i.e., it can launch a coordinated attack. 
In addition, the attacker can have access to a sample of historian data logs with a large number of DPs in it. 
It has enough computing capability to execute causal discovery algorithm on a larger set of data. 
A dataset considered to be large when the number of features in it is larger, and not necessarily the number of records be large. A large number of records can be expected to produce a \emph{better} causal graph, which we are not concerned in this paper.


\subsection{Defender Model}
Defender in this case is aiming to reduce the information leakage via historian data logs. 
Because the historian data log of a CPS is time-correlated readings of sensors and actuators, the correlation or the dependency among the DPs are not possible to eliminate completely (our work in this paper can be taken as evidence for such a claim). 
In one extreme, a CPS may decide to completely stop the disclosure of any historian data logs and in the other extreme, disclose all data logs after (anonymizing) the DPs. 
We consider a realistic scenario, the system defender release a portion of historian data log without complete anonymization in order to avail advanced analytics like the one we propose in this paper.

\section{Constructing Causal Graphs in CPS}
\label{sec:causal-graph-cps}
Causation has been defined long back (1956) by Reichenbach as follows:

"If two random variables X and Y are statistically not independent, then there is one of three possibilities: (a) X causes Y, (b) Y causes X, or (c ) there exists a third variable Z that causes both X and Y". 

Possibility (c) is significantly different from the possibilities (a) and (b) in the sense that there exists a third variable $Z$ that causes both $X$ and $Y$. 
For complete discovery of causal relation, the presence of this third random variable $Z$ in the data set is important. 
While the possibilities (a) and (b) indicates a cascading causal dependency, the third possibility can produce two distinct causal relations among the variables: i) common parent and ii) common child. 
In this section, we shall make use of this definition to both define causal graph and 
verify a causal graph using data from a CPS.

\subsection{Basics in Causal graphs and CPS}
A causal graph is represented by a directed acyclic graph $G=(V,E)$ with a set $V$ of $n$ nodes and a set $E$ of directed edges of the form $(v_i,v_j)$, where $v_i \in V$ has causal affect on $v_j\in V$. 
When a CPS is represented using a causal graph, each node $v_i\in V$ in the graph represents a DP. 
Alternately, a historian that collects data from a CPS considers a DP as a feature and a set of DPs is represented by a features vector $\vec{F}$. 
At each time step $t$, the readings from this set of DPs is represented by a feature vector $\vec{F}_t$ which is then stored in a data log forming a record at that timestep. 
Hence, $i^{th}$ value in the feature vector $\vec{F}_t$ indicates the value $f_i^t$ of $i^{th}$ feature at time step $t$.
For instance, a feature vector 
$$\vec{F} = (P101, P102, LIT101, MV101, FIT101)$$ indicates the set of DPs in stage-1 of SWaT, i.e., the historian data log contains a data set corresponding to the DPs in stage-1. 
Thus, the set $V$ of nodes in the corresponding causal graphs $G$ has a one-to-one correspondence with the set of DPs in Stage-1.
We use a node $v_i$ in the causal graph, a DP in a CPS, a feature $f_i$ in the feature vector $\vec{F}$ in the historian and a random variable in conditional probability in a causal graph to indicate a same thing, because we can have one-to-one mapping among these symbols. 

An edge $(v_i,v_j)$ in the causal graph $G$ indicating $v_j$ dependent on $v_i$ corresponds to a conditional probability $Pr(v_j|v_i)$, i.e., the probability of $v_j$ given $v_i$. 
At this point, we do not bother about whether a feature corresponding to $v_i$ represents a continuous or discrete random variable. 
Thus, an edge $LIT101 \dashrightarrow MV101$ in a causal graph $G$ indicates the conditional probability $Pr(MV101 | LIT101)$ and this probability is considered as the parameter of this edge. 
Given a causal graph having such edges, one can estimate the parameters using a historian data log.
Similarly, a conditional probability $Pr(P101 | LIT101, FIT101)$ is presented by two edges, $LIT101 \dashrightarrow P101$ and $FIT101 \dashrightarrow P101$, in the graph $G$, i.e.,

\begin{equation}
\begin{aligned}[c]
Pr(P101 | LIT101, FIT101) \\
\end{aligned}
\qquad\Longleftrightarrow\qquad
\begin{aligned}[c]
LIT101 \dashrightarrow P101 \\ 
FIT101 \dashrightarrow P101 \\
\end{aligned}
\end{equation}

Extending this concept, a complete CPS can be thought of as a joint probability distribution $Pr(F)$, where $F$ indicates an unordered list of DPs of the CPS under consideration for which we aim to construct or learn the causal graph. 
Such an arrangement can help us to slice the whole CPS into smaller subsystems and we can focus on a specific portion of the CPS to perform our analysis. 
The joint probability distribution $Pr(F)$ can be expressed in term of the product of a set of conditional probabilities, 
\begin{equation}
Pr(x_1,x_2, ..., x_n) =  Pr(x_1)\prod_{i=2}^{n}  Pr(x_i | Pa(x_i)
\label{eq-pr-F}
\end{equation}

Each term in the product at the right hand side of  Equation \ref{eq-pr-F} is a conditional probability, except the last term, i.e., $Pr(x_1)$. 
A graphical representation of each of the factors on the right hand side essentially displays a Bayesian Network that is used in causal models for the display of the causal relationship between a pair of random variables.
The causal graph corresponding to this joint distribution $Pr(F)$ need not always represent an exact causal structure of a CPS, 
where every node has a causal relation with any other node in the graph. 

Let us consider a causal graph in Stage-1 of SWaT that can be constructed based on domain knowledge, i.e., the causal domain graph, as shown in Figure \ref{fig:swat-stage1-data-knowledge}(a). 
In this causal graph, we have a restricted number of edges, i.e., the graph is not a completer graph, nor every node is causally dependent on all other nodes. 
We can represent this causal graph in Stage-1 using a joint probability distribution as follows.

\begin{equation}
\begin{aligned}[c]
Pr(P101, P102, LIT101, MV101, FIT101) 
\end{aligned}
 = 
\begin{aligned}[c]
Pr(P101|LIT101)  Pr(P102|LIT101) Pr(MV101|LIT101) Pr(FIT101|) \\ Pr(LIT101) 
\end{aligned}
\end{equation}

Note that the number of product terms in right hand side is equal to the number of edges in the causal graph. 
Further, given such a causal graph structure, we can estimate the parameters for each of the edges in the graph using a corresponding data set, like historian data logs in Stage-1 in SWaT. 
Hence, such a graphical causal graph can be validated using the estimated values of the parameters only if the graph has no cyclic dependency among the variables represented in the graph, i.e., the graph need to be an acyclic directed graph. 

\subsection{Rules for Constructing Domain Graphs in CPS}
We propose two rules to construct a causal domain graph corresponding to a CPS by repeated application of the rules. 
Note that we are not interested to estimate the parameters values of the edges in the graph at this point. 
Construction the causal domain graphs require an extensive amount of knowledge of the control logic used in a CPS, i.e., each of the Programmable Logic Controllers (PLCs) in the CPS, and the physical coupling of the hardware components in it. 
Essentially, a significant amount of manual effort is needed to build the causal domain graphs, compared to that for learning the causal graphs from operational data logs.


A $(DP_i, DP_j)$ edge in the graph indicates that the value of $DP_j$ is dependent on the value of $DP_i$. 
This dependency can be due to either the control of the CPS or a physical coupling.
Hence, we use the basics of CPS design principles, as:
\begin{itemize}
    \item \textbf{Control Dependency:} Control logic to operate a set of DPs implanted in a PLC that receives the sensed values from a set of sensors and sends actuation commands to a set of actuators. To derive the dependency of $DP_j$ on $DP_i$, we do not use the knowledge of the exact control program running on a PLC, rather we use the knowledge of the functional requirements that associate the two DPs $DP_i$ and $DP_j$. 
    Therefore, if there exists a functional requirement such that a set of DPs is used as input (say, input DPs) and a set of DPs as output (say output DPs), then we draw a dependency edge from each DP in the input set to each DP in the output set.

    \item \textbf{Physical Coupling:} Specific to the cases of CPS design, the coupling of physical components, rather than control programs, play an important role in the overall operation of the CPS. For instance, if a motorised valve placed on a water pipe is $Open$ then a flow sensor attached to that pipe should automatically show a non-zero flow value in normal operating condition. 
    Conversely, if a flow sensor attached to a pipe shows a non-zero flow then the motorised valve attached to that pipe must be $Open$ in normal operating condition. 
    Such a physical coupling need not be captured in a functional requirement, rather it is the requirement of close monitoring of the correct operation of the plant. 
    In general, by observing the physical structure of the plant and combining the operational requirement of monitoring the plant status, we identify a dependency between a pair of DPs, which we call as physical coupling. 
    We represent such physical coupling as an edge from an actuator to a sensor, and not the other way around.
\end{itemize}

We consider three reasons for constructing the causal domain graphs in a CPS. 
Note that a causal domain graph is essentially a model of a CPS that can be used for advanced analytics like the one we present in this paper.
First, we want to compare structure of these graphs with the ones computed out of historian data logs, i.e., the causal learnt graphs. 
Second, we want to discover the distribution of the parameters of these graphs that can be estimated using historian data logs. 
The knowledge of such parameters can then be used in effective causal inference. 
Third, the causal domain graphs can be used to identify the impact of a cyber attack on a set of DPs even when a CPS is not operational.

\subsection{Verifying Causal Graphs using CPS Data}
\begin{figure}[hbt!]
  \centering
  \includegraphics[width=0.9\textwidth]{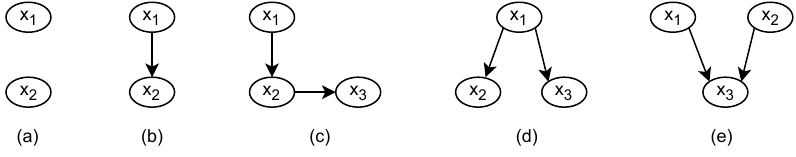}
  \caption{ Different cases of causal graphs}
  \label{fig:math-causal-graph}
\end{figure}

Informally, if a random variable $Z$ is caused by $X$, then the hypothesis that a change in $X$ will cause a change in $Z$, i.e., keeping $X$ constant will not cause a change in $Z$, holds true.  
Because causal graphs rely on conditional (in)dependence, we shall explore the statistical formulation of conditional (in)dependence along with Bayes theorem for the estimation of the parameters in a causal graph using historian data logs. 
We shall look at four different cases of causal graphs as shown in Figure \ref{fig:math-causal-graph}. 
We assume that a causal graph can be decomposed into a combination of subgraphs of the forms shown in Figure \ref{fig:math-causal-graph}.

We assume that each of DP representing a random variable assumes a set of discrete values. 
Such an assumptions can be supported by the fact that every sensor can cause a change in the state of a set of actuators when it crosses a threshold. 
Note that both the change in the state of an actuator or a change in the interval of the values of a sensor are examples of events in a CPS, that we consider in this paper. 
Only when a sensor crosses the boundary of an interval, it can cause a change in the states of a set of actuators via control logic executed in a PLC (programmable Logic Controller) that connects the sensors and the actuators and other PLCs in a CPS.
For instance, in a control logic, when LIT101 that measures water level in tank T101 indicates \emph{Low} (Low is an interval of the values of LIT101),  MV101 should be opened to allow water to flow into the tank. 
Similarly, when LIT101 indicates \emph{High} (High is an interval of the values of LIT101, different from that of Low), MV101 must be closed to avoid overflow of water in the tank. 
Further, when LIT101 is not Low, then the pump P101 can be switched on, otherwise none of P101 and P102 can be switched on. 
Thus, the range of values of LIT101 can be divided into three intervals, namely: \emph{Low}, \emph{Medium} and \emph{High}. 
So, we propose to consider all the DPs to have discrete values, i.e., a DP representing an actuator can already be considered as a discrete RV and a DP representing a sensor needs to be converted into a discrete RV with the help of a certain number of intervals of the values of the sensor. 
Note that the intervals of the values of a sensor can be created based on domain knowledge on the operating principles or by histogram analysis of the values of DPs. 

\subsubsection{Simple Statistical Independence}
Let us consider two RVs, $X_i$ and $X_j$, corresponding to two DPs in a CPS. 
A causal graph using $X_i$ and $X_j$ can be represented in two ways: i) the variables are not dependent on each other, i.e., they are statistically independent, and ii) $X_j$ is dependent on $X_i$. 
In either of the cases, we explore the concept of joint  distribution using $X_i$ and $X_j$. 
Assume that $X_i$ and $X_j$ have $n_p$ and $n_q$ distinct values respectively. 
Let us use the symbols $=$ and $==$ to indicate an assignment and a check of equality respectively. 
We say that $X_i$ and $X_j$ are independent if following holds for any pair of their values $(x_m^i, x_n^j)$:
\begin{equation}
Pr(X_i = x_m^i, X_j=x_n^j) == Pr(X_i = x_m^i)Pr(X_j = x_m^j)
\end{equation}

Alternately, $X_i$ and $X_j$ are independent if for any pair of values of $X_i$ and $X_j$: 
\begin{equation}
Pr(X_i = x_m^i| X_j=x_n^j) == Pr(X_i = x_m^i) \hspace{0.5in} \mbox{ or } \hspace{0.5in} Pr(X_i = x_m^i| X_j=x_n^j) == Pr(X_j = x_n^j)
\end{equation}

Figure \ref{fig:math-causal-graph}(a) shows a causal graph of $X_i$ and $X_j$, where the variables are independent of each other. 
For example, let us consider two DPs P101 and FIT101 in Stage-1 in SWaT.  
P101 being an actuator has two states (On and Off) and FIT101 being a sensor can have two states (Low and High) after creating the intervals of its values. 
In order to test the independence between P101 and FIT101, we need to check the condition for each of four pairs of values: (P101 = On, FIT101 = Low), (P101 = On, FIT101 = High), (P101 = Off, FIT101 = Low) and (P101 = Off, FIT101 = High). 
In a joint distribution $Pr(P101,FIT101)$, $Pr(FIT101)$ or $Pr(P101)$ can be obtained by using marginal distribution of FIT101 or P101 respectively.  

\subsubsection{Simple Statistical Dependence}
A causal graph with two random variables, $X_i$ and $X_j$, where $X_j$ is dependent on $X_i$ is shown in Figure \ref{fig:math-causal-graph}(b), where $X_j = x_2$ and $X_i=x_1$. 
The causal graph can be represented by a joint distribution 
\begin{equation}
Pr(X_i,X_j) = Pr(X_j|X_i)Pr(X_i)
\end{equation}

The probability $Pr(x_1)$ is considered as prior probability and any posterior probability can be computed by applying Bayes theorem.  
Let us consider two DPs, MV101 and LIT101. 
Table \ref{tab:mv101-lit101} shows a possible combination of the values of MV101 and LIT101 with their respective probabilities. 

\begin{table}[h]
  \begin{tabular}{ |c | c | c | c |}
    \hline
     & LIT101 = Low  & LIT101 = Medium & LIT101 = High \\ \hline
    MV101 = Open  & $p_{11}$ & $p_{12}$ & $p_{13}$ \\ \hline
    MV101 = Close & $p_{21}$ & $p_{22}$ & $p_{23}$ \\
    \hline
  \end{tabular}
   \caption{Joint probability for each pair of values of MV101 and LIT101.}
  \label{tab:mv101-lit101}
\end{table}

\subsubsection{Conditional Independence and Joint Probability Distribution}
Conditional probability is extended to conditional independence. 
Statistically, we know if $Pr(A|B) = Pr(A)$ then the two events $A$ and $B$ are simply independent (Figure \ref{fig:math-causal-graph}(a)).
In conditional independence, we involve three events where the premise is that two events $A$ and $B$ are independent given another event $C$ with $Pr(C)>0$, and the probability is denoted by $Pr(A,B|C)$.
If $A$ and $B$ are conditionally independent given another event $C$, then 
\begin{equation}
Pr(A\cap B|C) = Pr(A|C) Pr(B|C) \hspace{0.5in} \mbox{ and } \hspace{0.5in} Pr(A|B,C) = Pr(A|C)
\end{equation}

In the domain of causal models, conditional independence is denoted by $A\Perp B|C$ to indicate two events A and B are conditionally independent given another event C. 
Alternately, $Pr(A\cap B|C) \neq Pr(A|C) Pr(B|C)$ implies that the two events $A$ and $B$ are not conditionally independent given another event $C$. 
The concept of conditional independence on events is extended to random variables. 
Two discrete random variables $X_i$ and $X_j$ are conditionally independent given another discrete random variable $X_k$ if for any combination of their values $x_p^i$, $x_q^j$ and $x_r^k$, the following holds:
\begin{equation}
Pr(X_i = x_p^i, X_j = x_q^j| X_k = x_r^k) = Pr(X_i = x_p^i|X_k = x_r^k) Pr(X_j = x_q^j|X_k = x_r^k)
\end{equation}

\subsubsection{Conditional Dependence and Joint Probability Distribution}
Figure \ref{fig:math-causal-graph}(e) shows a graphical representation of conditional dependence among three random variables. 
Mathematically, it is represented as 
\begin{equation}
Pr(x_3|x_2,x_1) = Pr(x_3|x_2) Pr(x_3|x_1) 
\end{equation}
For instance, if P101 is On, then probability of MV101 is Open is less likely than LIT101 is Low. 
This is evident from the control logic in Stage-1 that MV101 is Open when LIT101 is not High.  

\subsection{Parameter Estimation in Causal Graphs}
Parameter estimation process requires as input a causal graph structure, i.e., a skeleton graph, and a corresponding data set. 
It is expected to produce two sets of probabilities. 
First, a set of prior probabilities where each probability is corresponding to each of those variables for which there is no in-coming edge in the causal graph. 
Second, a set of conditional probabilities where each probability is for each of those nodes where there is at least one in-coming edge in the causal graph. 
We use two measures for the estimation of both the probabilities:  
Maximum Likelihood Estimation (MLE) 
and 
Bayesian Estimation (BE). 

\subsubsection{MLE} 
Informally, MLE is one of the methods for estimating the parameters of an unknown distribution that have been used to generate a data set, i.e., MLE takes a data set $\mathbf{x}$ and a parameter $p$ that is assumed to be used in the unknown distribution and produces $\hat{p}$ that is an estimation of the parameter of the unknown distribution that maximizes the probability $Pr(\mathbf{x}|\hat{p})$. 
Note that MLE can estimate the parameter only for those distributions that uses only one parameter, like Bernoulli, Binomial or Exponential distribution. 
Also, MLE estimates $Pr(\mathbf{x}|\hat{p})$ and not $Pr(\hat{p}|\mathbf{x})$, statistically these two are significantly different. 
In our case, we are interested to find MLE for each orientation of the edges in a causal graph and check if the estimated values are optimized for the causal graph. 

For example, consider Stage-1 in SWaT that has five random variables $P101, P102, LIT101, MV101$ and $FIT101$. 
Assume, at a particular time step $t$, these five variables indicates the values as On, Off, Not Low, Close and Low respectively. 
Assume that the distribution that produces the values has one parameter $p$. 
The likelihood function can be defined as the probability of observing the values given the parameter $p$, i.e., 

\begin{equation}
\begin{aligned}[c]
L(On, Off, Not Low, Close, Low; p) 
\end{aligned}
 = 
\begin{aligned}[c]
Pr(P101 = On, P102  = Off, LIT101= Not Low, MV101 = Close, \\ FIT101 = Low| p) 
\end{aligned}
\end{equation}

MLE of $p$, denoted by $\hat{p}$, is a value of $p$ that maximizes $L(On, Off, Not Low, Close, Low; p)$. 
In general, MLE is computed by equating the derivative of the probability distribution with respect to its parameter $p$. 
We do not delve deeper more in MLE as we shall use the formulation to compute the MLE values using certain tools. 

\subsubsection{BE} 
As an alternative to MLE, Bayesian Estimation (BE) is an estimation that minimizes the posterior expected value of a loss function $L(p,\hat{p})$, where $p$ is a parameter of a prior distribution. 
Informally, the idea is that before we observe any new data, we already have certain prior knowledge about the distribution that produces the data, and the knowledge can be used to estimate the parameter given the observed data. 
Bayesian estimation make use of Bayes theorem that has four components, where three components (prior $Pr(A)$, likelihood $Pr(B|A)$ and evidence $Pr(B)$) are used to compute the forth component (posterior $Pr(A|B)$). 

Consider a parametric model $Pr(x|\mathbf{p})$ for probability of $x$ with parameter $\mathbf{p}$, where $\mathbf{p}$ is a random variable (unlike that in MLE where $p$ denotes a particular value). 
Consider $Pr(\mathbf{p})$ as the prior distribution for the random variable $\mathbf{p}$. 
We aim to compute the distribution for the parameter $\mathbf{p}$ given the observed data $x$, i.e., $Pr(\mathbf{p}|x)$. 
Bayesian estimation of the parameter $\mathbf{p}$ is the one that minimizes $L(p,\hat{\mathbf{p}})$.
Like MLE, we do not delve deeper more on Bayesian estimation as we shall use the formulation to compute the BE values using certain tool. 

\subsection{Causal Inference in CPS}
Informally, causal inference is a process that given a causal graph and a set of observed values computes the posterior probabilities of the values of the other variables in the graph.  
A wide range of inference algorithms has been proposed, a survey can be found in \cite{LiuyiACMSurvey2021}. 
However, we are interested in simple inferences that can be applied on a relatively simple graph structures like \emph{chain} or \emph{fork}, \emph{collider} or \emph{confounder}.

Consider the causal graph with two variables shown in Figure \ref{fig:math-causal-graph}(b) where $x_1$ causally affects $x_2$, where the graph can be represented by a joint distribution $Pr(x_1,x_2) = Pr(x_2|x_1)Pr(x_1)$.  
Assume that a value of $x_2$ is observed and we want to infer the probability of $x_1$ having a specific value, i.e., we want to compute $Pr(x_1|x_2)$. 
In general, Bayes theorem can used for causal inference that computes posterior distribution using prior, likelihood and evidence from past data. 
Essentially, we can compute 
\begin{equation}
Pr(x_1|x_2) = \frac{Pr(x_2|x_1)Pr(x_1)}{Pr(x_2)}
\end{equation}
Note that $Pr(x_2)$ is the marginal probability of $x_2$ given the joint probability distribution $Pr(x_1,x_2)$. 

For instance, consider the domain causal graph in Stage-1 in SWaT (Figure\,\ref{fig:swat-stage1-data-knowledge}(a)).  
We can infer the probability of P101 being On given that LIT101 indicates Low, i.e., $Pr(P101 = On | LIT101=Low)$; such a situation indicates an anomalous behavior of the plant. 
The probability can be computed as 
\begin{equation}
Pr(P101 = On | LIT101 = Low) = \frac{Pr(LIT101 = Low |P101 = On)Pr(P101 = On)}{Pr(LIT101 = Low)}
\end{equation}
Inferring such a probability using a causal graph (constructed using historian data log for instance) can significantly help an attacker to select a target attack point in a CPS like SWaT or to identify the DPs that are impacted due to an attack launched on another set of DP for effective forensics. 


\section{Experimental Setup: Data, Algorithm and Tool for Causal Graphs}
In this section, we make use of some learning algorithms that, given a data set, produces a causal graph. 
More specifically, a number of causal graphs are generated and one among them is returned by the algorithm based on some score (the method to compute the score is also given as input to the learning method). 
Note that our aim is not to discover any new casual discovery algorithm in this paper. 
Broadly, there are three types of causal discovery algorithms: constraint-based, score-based and functional causal model-based.
For the purpose of this paper, we select two types of algorithms such that both rely on statistical tests of conditional (in)dependence: constraint-based and score-based. 

\subsection{Peter Clarke Algorithm}
A well known algorithm in the constraint-based category of causal graph learning algorihms is Peter-Clarke (PC) algorithm \cite{glymour2019review} that takes quantitative data as input. 
The Algorithm starts by assuming a Complete undirected graph, where each node represents a single variable and keeps on building the causal graph over several iterations by removing edges and then introducing the direction to the edges. 
The algorithm takes either a procedure that can compute statistical conditional independence of the form $A,B\Perp C$ or a method that can compute and compare fitting scores, e.g., Bayesian Information Criteria (BIC), before and after removing a directed edge. 

PC Algorithm takes a data set as input that satisfies four criterion to obtain a \emph{good} causal graph: 
\begin{itemize}
    \item Acyclicity - The causal relations among the variables can be represented by a Directed Acyclic Graph (DAG).
    \item Markov property - Each node is  independent of their non-descendants when conditioned on their parents. 
    \item Faithfulness - All true  conditional independencies are present in the data set.
    \item Sufficiency - For any pair of nodes, there is no common external cause, i.e., no missing confounder.
\end{itemize}

One of the short comings in PC algorithm is that it may not be able to provide direction to every single edge due to, for instance, if the number of variables are too large, or the data sample has missing confounder(s).

\subsection{Hill Climb Search Algorithm}
Hill Climb (HC) search algorithm is one of the heuristic based algorithm, where, in every iteration $i$, a solution $S_i$ is identified based on the available solutions without looking into a solution $S_j$ in any earlier iterations $j$, where $j<i$. 
The algorithm starts with a skeleton causal graph consisting of a set of variables as nodes and an empty edge set in the graph. 
For every node $x_i$, a set $PC_x$ of nodes containing potential parents and potential children is identified based on certain heuristic algorithm that takes as input a data set corresponding to the nodes in the graph and the node $x_i$. 
At each iteration $i$, a causal graph $G_i$ is formed by applying a combination of three operations: addEdge, removeEdge and reverseEdge. 
An edge of the form ($x_i,x_j$) can be added, i.e., addEdge operation can be applied,  only if $x_j$ is in $PC_{x_i}$. 
For a causal graph at an iteration.
A score is then computed based on one of the scoring methods, like Baysian Information Criteria (BIC).
The algorithm terminates when there is no change in the score in last $k$ iterations (details of the algorithm can be found in \cite{IoannisHC2006}).

\subsection{Chow-Liu (CL) Algorithm}
CL algorithm is relatively simple causal structure learning algorithm compared to the other two algorithms, HC and PC algorithms. 
CL algorithm takes as input a set data set and a node $v_r$ as root node and produces as output a causal graph rooted at $v_r$. 
In brief, the algorithm starts by assigning a weight $w_{ij}$ to each edge of the form \{$v_i,vj$\}, i.e., it starts by considering the graph as an undirected graph.
The algorithm work on discrete variables, each variable has a probability for each of its values.
An edge is considered to represent a joint distribution of two of its end points. 
Essentially, the weight $w_{ij}$ of an edge \{$v_i,vj$\} is a function of the probability distributions of the two nodes and the edge, as follows. 

\begin{equation}
w_{ij} = \sum_{(p,q)} Pr(x_i=x_p^i, x_j=x_q^j)\log \left[ \frac{Pr(x_i=x_p^i, x_j=x_q^j)}{Pr(x_i=x_p^i)Pr(x_j=x_q^j)}\right]
\end{equation}

It is assumed that two variables $x_i$ and $x_j$ corresponding to two nodes have certain numbers of discrete values and $Pr(x_i=x_p^i, x_j=x_q^j)$ indicates the probability of a particular pair of values of the variables, $x_p^i$ and $x_q^j$ respectively. 
Once the weights are assigned to each of the edges, the algorithm find a spanning tree rooted at $x_r$ such that the tree has highest weight across all the edges. 
Finally, the direction of the edges are assigned as traversing from the root node to the leaf nodes in the tree (further details can be found in \cite{GlymourCausalCL1991}). 

\subsection{{{bnlearn}} Package for Causal Discovery}
{bnlearn} (Bayesian Network Learning and Inference) \cite{bnlearnR2022} is an open-sourced   tool that is exclusively developed to working with graphical causal models. 
The package supports i) causal structure discovery from data, ii) estimation of parameter values given a data set and a causal graph, iii) functions for causal inference given a causal graph, and iv) some of the techniques for comparing the causal graphs (e.g., compare a causal domain graph and causal learnt graph). 
The package has been originally developed using R language and recently a Python package with same name, bnlearn \cite{bnlearnPython2022}, has been released.
We use Python3 implementation of bnlearn for the purpose of this work. 

We are interested in three types of functionalities of the bnlearn Python package: causal discovery, parameter estimation and inference. 
bnlearn supports all these functionalities when all the random variables in a data set are discrete random variables. 
We use three learning algorithms: constraint-based search (i.e., PC algorithm), score-based search (i.e., HC Search algorithm) and local discovery (i.e., CL algorithms). 
The package takes as input \emph{cs}, \emph{hc} and \emph{cl} to indicate PC, HC and CL algorithms respectively. 

Each of these algorithms takes as input a scoring method that is used as a stopping criteria for the algorithms.
The scoring method is supplied as a parameter by \emph{score}, which can be one of four methods: chi-square test (indicated by \emph{chi2}) for testing a null hypothesis on conditional independence, Bayesian Information Criteria (indicated by \emph{bic}), and K2 score (\emph{k2}), and Dirichlet posterior density score (\emph{bdeu}). 
Essentially, each of the structural learning algorithm can be used with any of these four scoring methods. 
Both the algorithm and the scoring method are provided as input to the \emph{structure\_learning.fit()} method in bnlearn package; \emph{hc} and \emph{bic} are defaults to the \emph{fit()} method for learning the causal structure.  
\emph{parameter\_learning.fit()} in bnlearn is used for estimating the parameters of a graph from data. 
Both MLE and BE estimators are supported by this method in bnlearn.


\subsection{Experimental Setup}
\label{sec:case-study-1-swat}
Historian data logs in SWaT are not completely anonymized, in the sense that the names of some of the random variables can be correlated to specific DPs in SWaT, but not all.
For instance, a DP with name MV101 can be associated with a particular motorised valve present in Stage-1 in SWaT, whereas a DP with AIT201 can be associated with a chemical sensor in Stage-2 in SWaT without indicating which exact chemical is sensed by this sensor. 
We have used about 1GB of such historian data logs in our work in this paper. 
Changing the size of historian sample may have some impact on a causal learnt graph.
We believe that such a change in the graph structure will not have any significant impact on the inferences that can be drawn from these graphs, and hence we do not explore more in this direction in this paper. 
%
%
We use the causal learnt graphs for causal inference in SWaT. 
In this case, we consider a set of known cyber attacks in SWaT and apply those in the learned graphs. 
Our aim in this part of the work is to demonstrate a usecase for causal graph in identification of Impacted DPs in each of the attacks in this paper. 

\subsection{Computing Environment}
We use a desktop computer with contemporary configuration like i5 with 8GB RAM. 
We use Python3 and related packages for the causal discovery, parameter estimation, and causal inference. 
In particular, Python3 packages like Pandas, Matplotlib and SciPy, in addition to bnlearn and pgmpy, are used in this work.

\section{Causal Graphs in SWaT}
\label{sec:causalgraphswat}
In this section, we build the causal domain graphs and the causal learnt graphs considering SWaT. 
Note that no historian data log is used for constructing the causal domain graphs and no domain knowledge is used for building the causal learnt graphs. 
In case of the causal domain graphs, we derive derive the causal relationships among the DPs from the control logic and physical coupling in SWaT.   
We consider each of the six stages SWaT separately in order to investigate stage specific dynamics in the causal graphs. 

\subsection{Causal Building Blocks in SWaT}
Construction of causal domain graphs requires extensive knowledge of the control logic of each of the stages, i.e., of each of the Programmable Logic Controllers (PLCs), and the physical coupling of the hardware components in that stage in the plant. 
Essentially, a significant amount of manual effort is needed to build the domain graphs, compared to that for the causal learnt graphs.
Recall that a DP is a physical device that is either a sensor or an actuator in the CPS, e.g., a pH sensor or a motorised valve.

A $(DP_i, DP_j)$ edge indicates that the value of $DP_j$ is dependent on the value of $DP_i$. 
Recall that such a dependency can be derived on the basis of the knowledge about two CPS design principles and hence we have two types of edges:
\begin{itemize}
    \item \textbf{Control Edge:}  
    An edge derived using control dependency and it is indicated by a dashed arrow as $DP_i \dashrightarrow DP_j$. 
    As stated earlier, control dependency can be derived from the control logic of a PLC or the functional requirement in a stage of SWaT.
    For instance, a motorised valve $MV101$ is $Open$ if water level in tank T101 indicated by a water level sensor $LIT101$ is not high, otherwise the valve be $Close$. 
    Thus, we represent such a control edge as $LIT101 \dashrightarrow MV101$ in the causal domain  graph.
    
    \item \textbf{Physical Edge:}  
    An edge constructed based on physical coupling and it is denoted by a solid arrow as $DP_i \rightarrow DP_j$, where $DP_i$ is an actuator and $DP_j$ is a sensor. 
    Though a pair of a sensor and an actuator can be physically close enough and cooperate, we consider that an actuator causally affects a sensor.
    For instance, if a motorised valve $MV101$ and a flow sensor $FIT101$ is attached in a close proximity to an input water pipe to a water tank, then the dependency is represented as $MV101 \rightarrow FIT101$ in the causal domain graph.
\end{itemize}

Constructing the causal domain graphs serves two purposes. 
First, we can perform a basic verification of the graphs by estimating their parameters using the historian data. 
Such a check can help to justify the strength of each of the edges and the nodes in the domain graph.
Second, the causal domain graphs can help to have a kind of cross check with the causal learnt graphs. 
Both the causal domain and learnt graphs can then be used with a high confidence to apply causal inference for advanced analytics like identifying impact of a cyber attack on the CPS. 
Note that identifying a set of DPs that can be targeted or impacted by a cyber attack may not be decided apriori.  
However, one can estimate the set of DPs that can be associated with an attack by providing the target of the attack. 
In the later section, we shall demonstrate how such analytics can be performed with the help of the causal learnt graphs. 

\subsection{Common Setup for Causal Domain and Learnt Graphs} 
To comprehend SWaT in respect of causal graph, we consider the CPS in phases to manage the complexity of the whole system, i.e., in each phase, we consider a limited number of DPs in SWaT. 
However, we do not restrict to single stage while applying the analytics of identifying the impacted DPs of a cyber attack (discussed in later section).
While executing the causal structure learning algorithms, we rely on the default parameters of each of the algorithms that we have used in this paper. 
For instance, PC algorithm that belongs to constraint-based learning algorithms uses parameters like alpha having a default value as 0.01 (as specified in the \emph{bnlearn} package).

\subsection{Causal Graphs in Stage-1: Feed raw water to SWaT}
Stage-1 in SWaT has five DPs (P101, P102, LIT101, MV101 and FIT101), out of which three are actuators (P101, P102 and MV101), i.e., each of these actuators represents a discrete random variable, and two are  sensors (LIT101, FIT101), i.e., each these sensors represents a continuous random variable.  
We descretise the continuous random variables to be used in the learning of causal structure. 
FIT101 values are divided into two intervals represented by two discrete values, Low and High, where Low indicates no flow and High indicates flow of water through the water pipe attached to it. 
Similarly, LIT101 values are divided into three intervals representing three discrete values, Low, Medium and High, where Low indicates a low water level, High indicates a high water level and Medium indicates a wter level that is neither high nor low in the tank T101. 
Note that discretisation of the sensor values can vary based on the knowledge of the designer or based on the requirement of the plant; consulting a design document can reduce its impact on the construction. 

\begin{figure}[h]
     \centering
     \begin{subfigure}[b]{0.23\textwidth}
         \centering
         \includegraphics[width=\textwidth]{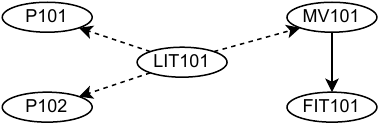}
         \caption{Causal domain graph.}
         \label{fig:y equals x}
     \end{subfigure}
     \hfill
     \begin{subfigure}[b]{0.23\textwidth}
         \centering
         \includegraphics[width=\textwidth]{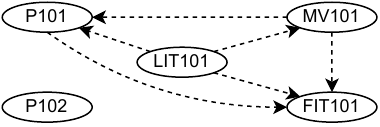}
         \caption{PC learnt graph.}
         \label{fig:three sin x}
     \end{subfigure}
     \hfill
     \begin{subfigure}[b]{0.23\textwidth}
         \centering
         \includegraphics[width=\textwidth]{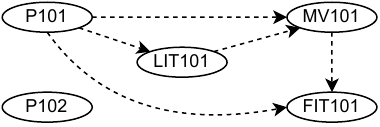}
         \caption{HC learnt graph.}
         \label{fig:five over x}
     \end{subfigure}
     \hfill
     \begin{subfigure}[b]{0.23\textwidth}
         \centering
         \includegraphics[width=\textwidth]{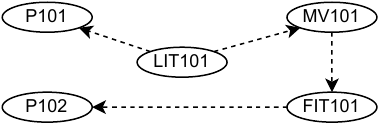}
         \caption{CL learnt graph.}
         \label{fig:five over x}
     \end{subfigure}
        \caption{Causal graphs in Stage-1 in SWaT. All the learnt graphs use BIC as scoring method.}
        \label{fig:swat-stage1-data-knowledge}
\end{figure}

\subsubsection{Causal Domain Graph in Stage-1:} Figure \ref{fig:swat-stage1-data-knowledge}(a) show the causal graph constructed using domain knowledge on the the operation of the Stage-1 in SWaT. 
A total of four edges can be drawn among five DPs in Stage-1, three of them are control dependency and one of them is due to physical coupling. 
The arc (LIT101, MV101) is created based on the logic that when LIT101 indicates Low water level in tank T101, MV101 must be opened to allow water flow into the tank and when LIT101 indicates High, MV101 must be closed to avoid overflow of water in the tank.
This is inline with the requirement of filling water tank in Stage-1.
Similarly, the arcs (LIT101, P101) and (LIT101, P102) are created based on the control logic that if LIT101 is not Low, then either P101 or P102 can be switched on to push water out of the tank; otherwise, none of the pumps be switched on to avoid underflow of water in the tank. 
This in inline with the requirement of pushing water from Stage-1 to Stage-2. 
MV101 and FIT101 are not controlling each other, but they are installed on the same water pipe and hence whenever water needs to flow thought the pipe, MV101 needs to be opened and FIT101 is used to sense the water flow rate. 
Alternately, if water flow needs to be stopped then MV101 be closed and hence FIT101 should indicate Low. 
Thus, the arc (MV101, FIT101) is due to physical coupling of the DPs in Stage-1. 

\subsubsection{Parameter Estimation for Stage-1 Causal Domain Graph:} Given the casual domain graph in stage-1, we estimate the parameters, based on historian data, for each of the nodes based on the incoming edge(s) for a simple verification of the graph.  
Note that we are not using the historian  data logs to build the graph, but to estimate the parameters of the graph. 
Because LIT101 is not having incoming arc, we simply find the estimated probability $Pr(LIT101=s_k)$, where $s_k$ indicates one of the three possible states of LIT101.  
Both MLE and BE estimators shows a similar result, i.e., LIT101 shows water level as Medium with highest probability (more than 0.7), Low is relatively unlikely (less than 0.05) and High is a non-negligible probability (about 0.2).  
\begin{figure}[h]
     \centering
     \begin{subfigure}[b]{0.23\textwidth}
         \centering
         \includegraphics[width=\textwidth]{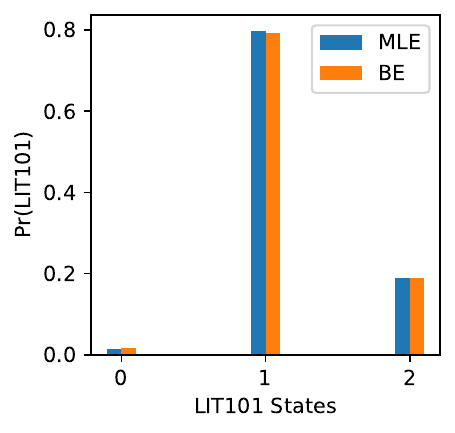}
         \caption{Parameter estimation:}
         \label{fig:y equals x}
     \end{subfigure}
     \hfill
     \begin{subfigure}[b]{0.23\textwidth}
         \centering
         \includegraphics[width=\textwidth]{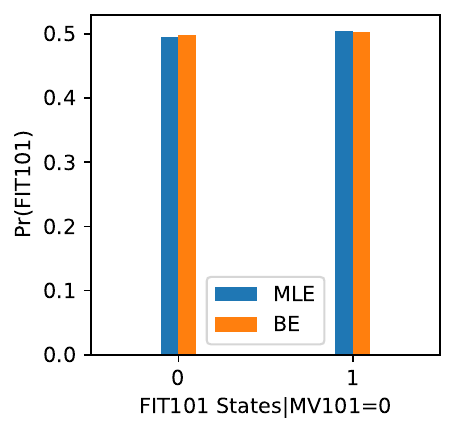}
         \caption{Parameter estimation:}
         \label{fig:three sin x}
     \end{subfigure}
     \hfill
     \begin{subfigure}[b]{0.23\textwidth}
         \centering
         \includegraphics[width=\textwidth]{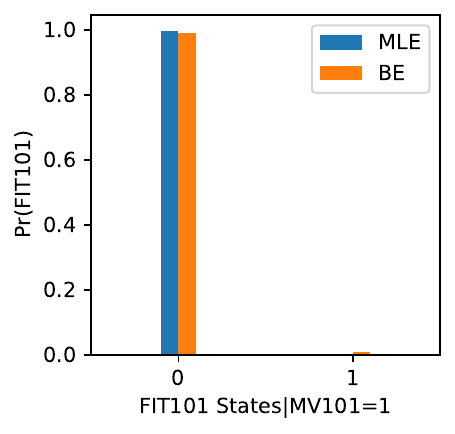}
         \caption{Parameter estimation: }
         \label{fig:five over x}
     \end{subfigure}
     \hfill
     \begin{subfigure}[b]{0.23\textwidth}
         \centering
         \includegraphics[width=\textwidth]{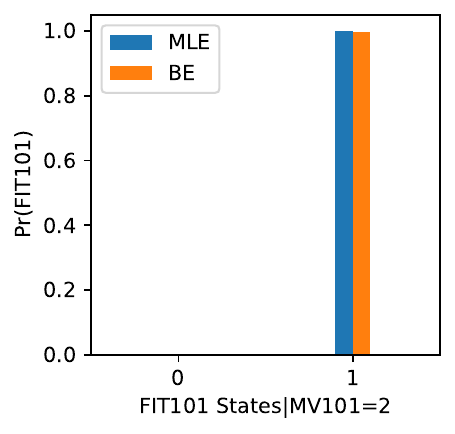}
         \caption{Parameter estimation: }
         \label{fig:five over x}
     \end{subfigure}
     \begin{subfigure}[b]{0.23\textwidth}
         \centering
         \includegraphics[width=\textwidth]{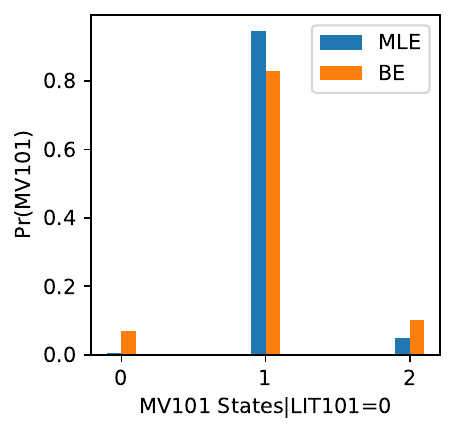}
         \caption{Parameter estimation:}
         \label{fig:y equals x}
     \end{subfigure}
     \hfill
     \begin{subfigure}[b]{0.23\textwidth}
         \centering
         \includegraphics[width=\textwidth]{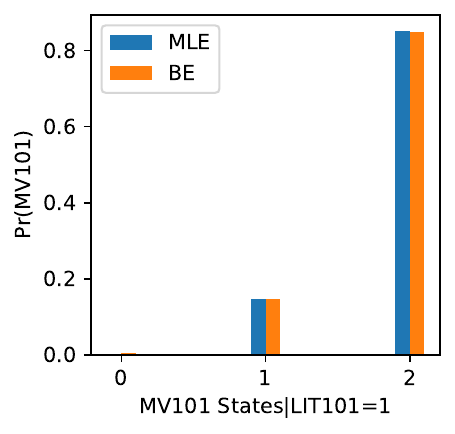}
         \caption{Parameter estimation:}
         \label{fig:three sin x}
     \end{subfigure}
     \hfill
     \begin{subfigure}[b]{0.23\textwidth}
         \centering
         \includegraphics[width=\textwidth]{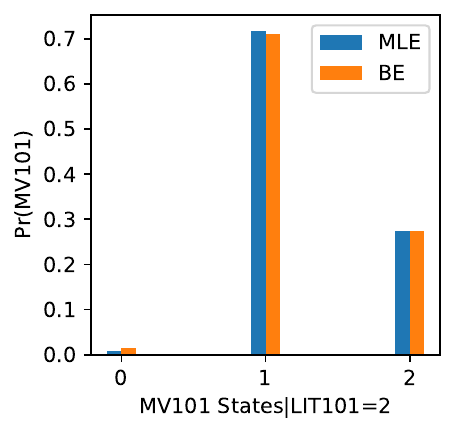}
         \caption{Parameter estimation: }
         \label{fig:five over x}
     \end{subfigure}
     \hfill
     \begin{subfigure}[b]{0.23\textwidth}
         \centering
         \includegraphics[width=\textwidth]{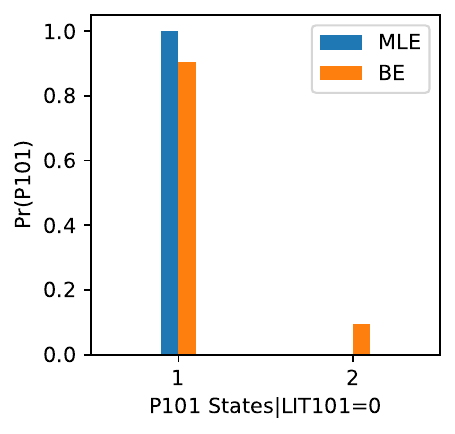}
         \caption{Parameter estimation: }
         \label{fig:five over x}
     \end{subfigure}
     \begin{subfigure}[b]{0.23\textwidth}
         \centering
         \includegraphics[width=\textwidth]{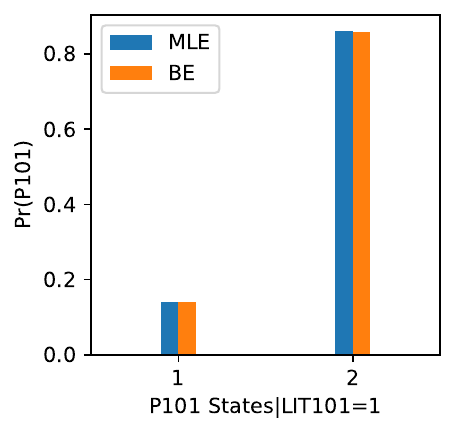}
         \caption{Parameter estimation: }
         \label{fig:five over x}
     \end{subfigure}
     \hfill
     \begin{subfigure}[b]{0.23\textwidth}
         \centering
         \includegraphics[width=\textwidth]{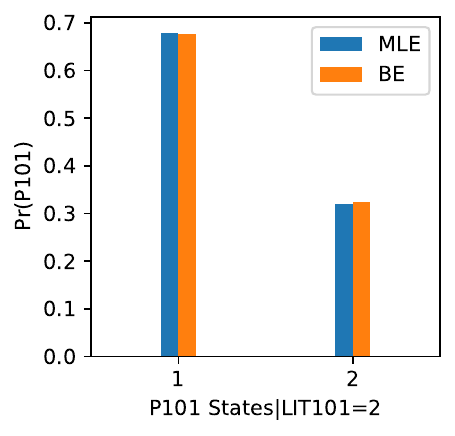}
         \caption{Parameter estimation: }
         \label{fig:five over x}
     \end{subfigure}
     \hfill
     \begin{subfigure}[b]{0.23\textwidth}
         \centering
         \includegraphics[width=\textwidth]{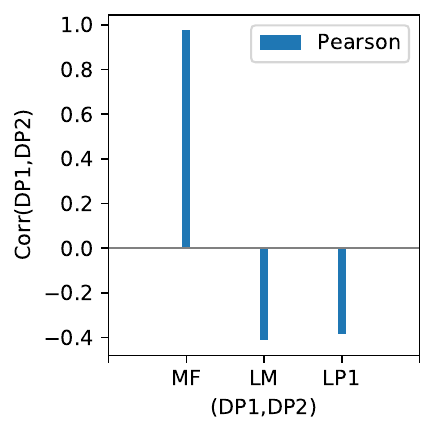}
         \caption{Corr. for pairs with edge }
         \label{fig:five over x}
     \end{subfigure}
     \hfill
     \begin{subfigure}[b]{0.23\textwidth}
         \centering
         \includegraphics[width=\textwidth]{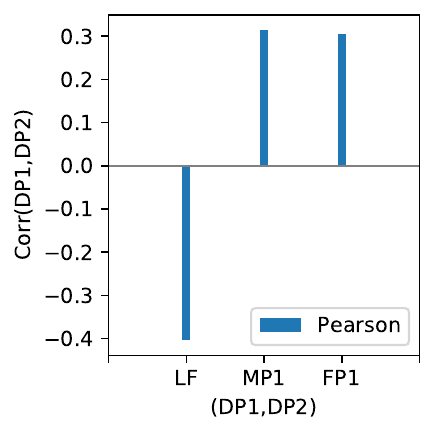}
         \caption{Corr. for pairs with no edge }
         \label{fig:five over x}
     \end{subfigure}
        \caption{Parameter estimation in causal domain graph in Stage-1 in SWaT. \emph{Corr.} indicates Pearson correlation. }
        \label{fig:swat-stage1-param-estimate-knowledge}
\end{figure}

All other nodes have a dependency on other node(s) in the graph, i.e., there is at least one incoming edge to it, let us investigate each of these cases. 
$Pr(MV101 | LIT101 = 0)$ for each of the states of MV101 is shown in Figure \ref{fig:swat-stage1-param-estimate-knowledge}(e). 
The historian data indicates MV101 has three states and it is most probable (>0.8) to be in close state when LIT101 is Low. 
The MLE estimate is relatively higher than that using BE. 
Alternately, given LIT101 is High, MV101 is Close (i.e., MV101=1) with a probability almost 0.7, whereas MV101 is Open (i.e., MV101 = 2) with a probability of more than 0.25 (Figure \ref{fig:swat-stage1-param-estimate-knowledge}(g)). 
If LIT101 is Medium, MV101 is Open with a probability more than 0.8, and this indicates that water keep in-flowing to T101 until LIT101 become High (Figure \ref{fig:swat-stage1-param-estimate-knowledge}(f)).
Thus, we can conclude that LIT101 indeed has an impact on MV101 and it is indicated the arc from LIT101 to MV101 in the causal graph. 
It will be interesting to see if a causal graph generated by using historian data has an arc between there two DPs, which is done later in this section. 

The pump P101 has two states, On (P101 = 2) and Off (P101 = 1).
It is Off given LIT101 indicates Low with a probability of almost 1.0 using MLE estimation (Figure \ref{fig:swat-stage1-param-estimate-knowledge}(h)).
BE estimation in this case is slightly lower (about 0.9) compared to MLE estimation; vis-a-vis, P101 is On with a probability about 0.1. 
Given LIT101 is High, P101 is mostly Off (probability > 0.65) and it is On with a relatively lower probability (about 0.3) (Figure \ref{fig:swat-stage1-param-estimate-knowledge}(j)). 
When LIT101 is Medium, P101 is switched on with probability greater than 0.8 and it is Off with a probability less than 0.2 (Figure \ref{fig:swat-stage1-param-estimate-knowledge}(i)).
Altogether, this indicates that it is not necessary that only LIT101 causes P101 to operate, but LIT101 has a causal affect on P101.
Hence, we can validate the arc from LIT101 to P101 with the support of estimated conditional probabilities. 

FIT101 can have two states, Low (FIT101=0) or High (FIT101 = 1), to indicate either water is not flowing or flowing respectively through the pipe attached to it. 
FIT101 can be in both the states, one at a time, given MV101 is 0 (a possible interpretation of this state of MV101 is that it is in inactive state or closed state) with almost equal probability (about 0.49 each) (Figure \ref{fig:swat-stage1-param-estimate-knowledge}(b)). 
However, according to operational logic in Stage-1, MV101 in state 0 is not very relevant for the correct functioning of the plant and therefore we shall focus more on the other two states of MV101. 
Given MV101 is Close, FIT101 shows Low with probability almost 1.0 (Figure \ref{fig:swat-stage1-param-estimate-knowledge}(c)) and it is High given MV101 is Open with a probability of about 1.0 (Figure \ref{fig:swat-stage1-param-estimate-knowledge}(d)). 
Hence, we conclude that historian data log supports the arc from MV101 to FIT101 (note that this arc is due to physical coupling). 
Further, we have looked at the correlation of the each of the pairs of the DPs in Stage-1. 
It turns out that Pearson correlation between MV101 and FIT101 is 0.8, which indicates that MV101 and FIT101 are positively correlated, i.e., either a higher value can be seen for each of these DPs or a lower value. 
In other words, if MV101 = 2 then FIT101 = 1, or if MV101 = 1 then FIT101 = 0. 
It is our choice to consider the arc direction as from MV101 to FIT101, representing a physical coupling.

\subsubsection{Causal Learnt Graphs in Stage-1:}
We use historian data logs to learn the causal graphs using three algorithms: PC, HC and CL; all using BIC as a scoring method. 
Note that it may not be easy to validate the learned graphs due to lack ground truth. 
However, we can cross check the similarity or dissimilarity (which is not necessarily be based on structural elements, like edges) between the causal domain and learnt graphs. 

Figure \ref{fig:swat-stage1-data-knowledge}(b) shows the causal learnt graph using PC algorithm, call it as PC learnt graph. 
The PC learnt graph is able to replicate three arcs (LIT101,P101), (LIT101, MV101) and (MV101, FIT101) as that of the causal domain graph (Figure \ref{fig:swat-stage1-data-knowledge}(a)) in Stage-1.
Interestingly, PC algorithm discovers three not-seen-before arcs: (P101, FIT101), (LIT101, FIT101) and (MV101, P101). 
The arc (LIT101, FIT101) is actually implied by the domain graph. 
LIT101 turns out to be a confounder for MV101 and FIT101, i.e., both MV101 as a treatment and FIT101 as an outcome of that treatment are causally affected by LIT101. 
The other two arcs cannot be implied by the domain graph. 
We speculate that these two arcs are due to external control logic running in PLCs across different stages or due to an implied control logic in the PLC in Stage-1. 
For instance, given LIT101 indicating Medium, P101 can push water to Stage-2 from T101 and MV101 can be opened to allow a continuous flow of water into T101. 
However, if MV101 is Open due to low water level in T101, then P101 cannot be switched on, and hence a dependency is revealed in the data set. 
A similar argument can be applied in case of the arc (P101, FIT101). 
This is an example of a case where the values of a DP (i.e., P101) is manifested as causally related to the values of another DP (i.e., FIT101), but our analysis of control dependency could not reveal it.

Figure \ref{fig:swat-stage1-data-knowledge}(c) shows the causal graph learned using HC algorithm, call as HC learnt graph. 
This algorithm could find two arcs as that in the domain graph, namely (LIT101, MV101) and (MV101, FIT101). 
Unlike LIT101 in PC algorithm, P101 has become a confounder for MV101 and FIT101 in this case. 
Also, the arc between P101 and LIT101 is reversed compared to both the causal domain graph and the PC learnt graph. 
Note that we have used a same data set to both the algorithms to learn the causal graphs.  
Any difference between the learnt graphs crops up due to the differences in the exploration in the causal relationships.
For instance, HC algorithm relies on identifying the patents and the children of node in the graph in the first place, whereas PC algorithm look for conditional independence to remove the edges from an undirected graph and then impose the direction of dependency to the edges.

Figure \ref{fig:swat-stage1-data-knowledge}(d) shows the causal graph learned using CL algorithm, call it as CL learnt graph. 
This graph is significantly different from those of the other three graphs, i.e., the causal domain graph, the PC learnt graph and the HC learnt graph. 
Three out four arcs are reversed compared to that in the domain graph. 
The arc between FIT101 and P101 is reversed as opposed to the PC learnt graph  and the HC learnt graph. 
CL algorithm is fundamentally a different algorithm that makes use of a user provided root node for learning a rooted directed acyclic graph as a causal graph.
Overall, while the graphs learned using different algorithms are not very similar in terms of their structure, it is not necessary that the graphs leads to a significantly different set of the causal inferences.
We shall explore causal inferences using these graphs while characterising cyber attacks on SWaT in next section.

\subsection{Casual Graphs in Stage-2: Chemical Dosing}
Eight DPs in Stage-2 of SWaT are used for data collection in the historian.  
Four of them are continuous RVs, that we have discretized by using expert knowledge. 

\subsubsection{Causal Domain Graph in Stage-2:} 
\begin{figure}[h]
     \centering
     \begin{subfigure}[b]{0.23\textwidth}
         \centering
         \includegraphics[width=\textwidth]{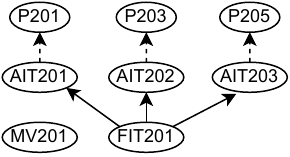}
         \caption{Causal domain graph.}
         \label{fig:y equals x}
     \end{subfigure}
     \hfill
     \begin{subfigure}[b]{0.23\textwidth}
         \centering
         \includegraphics[width=\textwidth]{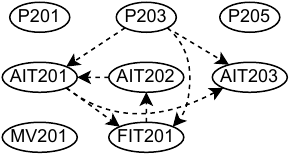}
         \caption{PC learnt graph.}
         \label{fig:three sin x}
     \end{subfigure}
     \hfill
     \begin{subfigure}[b]{0.23\textwidth}
         \centering
         \includegraphics[width=\textwidth]{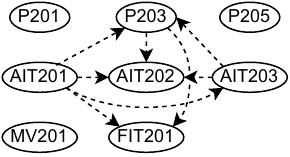}
         \caption{HC learnt graph.}
         \label{fig:five over x}
     \end{subfigure}
     \hfill
     \begin{subfigure}[b]{0.23\textwidth}
         \centering
         \includegraphics[width=\textwidth]{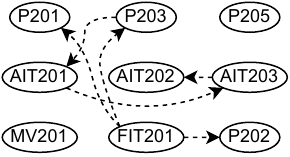}
         \caption{CL learnt graph.}
         \label{fig:five over x}
     \end{subfigure}
        \caption{Causal graphs in Stage-2 in SWaT. All the learnt graphs use BIC as scoring method.}
        \label{fig:swat-stage2-data-knowledge}
\end{figure}

Figure \ref{fig:swat-stage2-data-knowledge}(a) show the causal domain graph in Stage-2. 
The sensing and actuating logic in Stage-2 is applied only when there is water flow as sensed by FIT201, so it is an independent RV in the domain  graph. 
Because all the chemical sensors are physically attached to the water pipe to which FIT01 is also attached, we have created a directed edge indicating physical coupling from FIT201 to each of these chemical sensors, i.e., AIT201, AIT202 and AIT203. 
Because a particular type of chemical is injected by switching on a corresponding pump only when the chemical level indicated by the chemical sensor is Low (per say), we have created a directed edge indicating control dependency from the chemical sensor to the corresponding pump, e.g., from AIT201 to P201. 
In this stage, the main water pipe is considered to be divided into two parts, one before the static mixer to which FIT201 is attached and the other after the static mixer to which MV201 is attached. 
Therefore, unlike in Stage-1, we do not considered FIT201 and MV201 have any physical coupling. 
Hence, MV201 remains isolated in the domain graph.  

\subsubsection{Parameter Estimation for Stage-2 Causal Domain Graph:} 
Like in Stage-1, we estimate the parameters of the domain graph. 
FIT201 being independent RV has the simple probabilities as $Pr(FIT201=0) \approx 0.22$ and $Pr(FIT201=1) \approx 0.74$ as estimated using MLE.
BE estimation of these parameters are similar ($\approx$ 0.23 and $\approx$ 0.74 for its states of 0 and 1 respectively). 
Because the parameters estimated using MLE and BE are almost similar in both Stage-1 and Stage-2, we shall discuss only the MLE estimated values for the parameters of the domain graph in the rest of the paper unless there is a significant difference in the estimated values by MLE and BE.

\begin{figure}[h]
     \centering
     \begin{subfigure}[b]{0.23\textwidth}
         \centering
         \includegraphics[width=\textwidth]{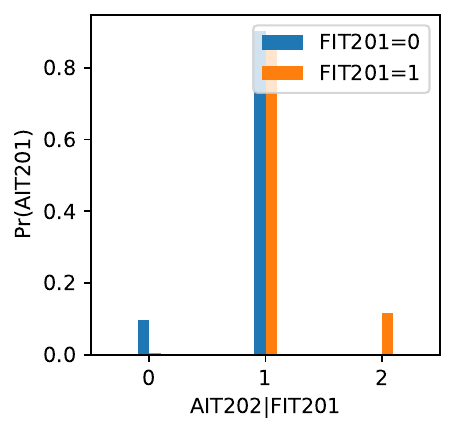}
         \caption{Parameter estimation for AIT201 given FIT201}
         \label{fig:y equals x}
     \end{subfigure}
     \hfill
     \begin{subfigure}[b]{0.23\textwidth}
         \centering
         \includegraphics[width=\textwidth]{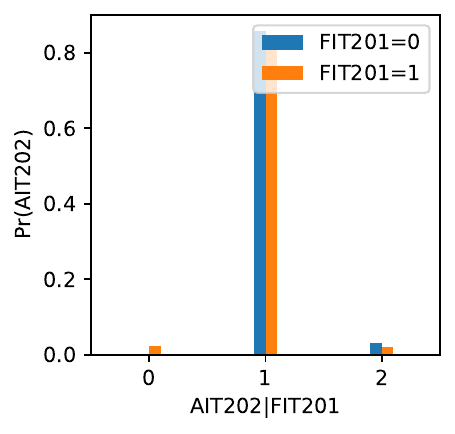}
         \caption{Parameter estimation for AIT202 given FIT201}
         \label{fig:three sin x}
     \end{subfigure}
     \hfill
     \begin{subfigure}[b]{0.23\textwidth}
         \centering
         \includegraphics[width=\textwidth]{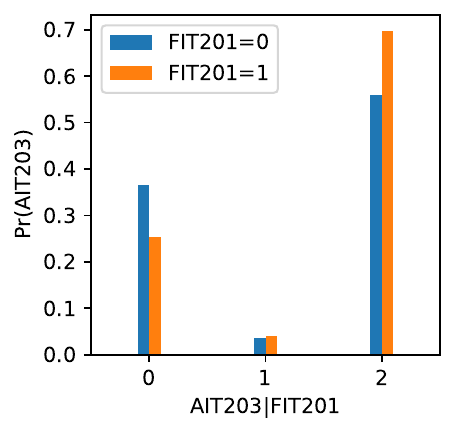}
         \caption{Parameter estimation for AIT203 given FIT201}
         \label{fig:five over x}
     \end{subfigure}
     \hfill
     \begin{subfigure}[b]{0.23\textwidth}
         \centering
         \includegraphics[width=\textwidth]{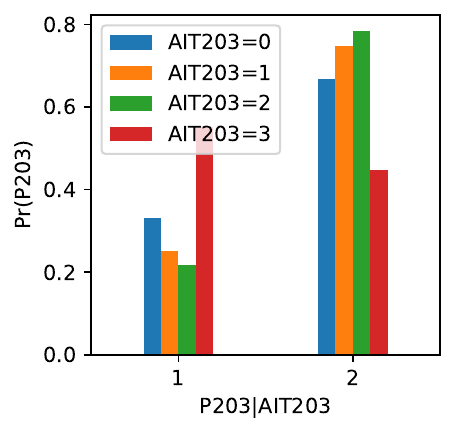}
         \caption{Parameter estimation for P203 given AIT203}
         \label{fig:five over x}
     \end{subfigure}
        \caption{Parameter estimation for the causal domain graph in Stage-2 in SWaT.}
        \label{fig:swat-stage2-param-estimate}
\end{figure}
The flow meter FIT201 causes the chemical sensors to change their values as depicted in Figure \ref{fig:swat-stage2-data-knowledge}(a). 
We estimate the probability given a particular state of FIT201. 
For instance, $Pr(AIT201=0|FIT201=0)\approx 0.1$ indicates that the probability of AIT201 is Low (i.e., AIT201 = 0) given FIT201 is Low is about 0.1. 
Further, one can validate the edge from FIT201 to AIT201 based on both  $Pr(AIT201=1|FIT201=0) \approx 0.8$ and $Pr(AIT201=1|FIT201=1) \approx 0.8$, (Figure \ref{fig:swat-stage2-param-estimate}(a)). 
Similarly, $Pr(AIT202 = 1 | FIT201=0) \approx 0.85$ or $Pr(AIT202 = 1 | FIT201=1) \approx 0.85$ can validate the edge from FIT201 to AIT202 (Figure \ref{fig:swat-stage2-param-estimate}(b)) 
Also, the edge from FIT201 to AIT203 can be validated based on the results shown in Figure \ref{fig:swat-stage2-param-estimate}(c). 

The pumps P201, P203 and P205 in Stage-2 is switched on or off based on the chemical levels indicated by the corresponding sensors. 
The estimated probabilities show that the pumps P201 and P203 are not switched on, and hence ignored from reporting the parameters.
Pump P205 is switched on and off for different states of the chemical sensor AIT203.
AIT203 has four states, indicated by 0, 1, 2 and 3, as can be derived from the distribution of its data and operation logic. 
Note that such a discretization of AIT203 need not be unique across designers and a change in the discretisation can have significant affect on the estimated parameters of a causal graph.
Figure \ref{fig:swat-stage2-param-estimate}(d) shows that, given any state of AIT203, pump P205 is both switched on and off with a significant probability. 
For instance, $Pr(P205=1|AIT203=3) \approx 0.6$ and $Pr(P205=2|AIT203=3) \approx 0.4$, where $P205=1$ indicate P205 is switched off when AIT203=3. 
Hence, we can validate the edge from AIT203 to P205 in the graph in Stage-2.

\subsubsection{Learnt Causal Graphs in Stage-2:} 
Figure \ref{fig:swat-stage2-data-knowledge}(b), Figure \ref{fig:swat-stage2-data-knowledge}(c) and Figure \ref{fig:swat-stage2-data-knowledge}(d) show the PC learnt, HC learnt and CL learnt graph respectively. 
In general, a large number of not-seen-before causal relations can be discovered by each of these algorithms using historian data.
PC algorithm learns only one edge (FIT201, AIT202) which is present in the corresponding domain graph. 
The edge between AIT201 and FIT201 is reversed in the learned graph. 
P203 turns out to be a confounder in this learned graph. 
Also, the graph shows that P203 has a causal effect on AIT201, AIT203 and FIT201. 
This is surprising because it shows that injecting chemical using P203 can alter the values of other chemical sensors than the one (i.e., AIT202) that exhort P203 to be switched on or off.  

The HC learnt graph (Figure \ref{fig:swat-stage2-data-knowledge}(c)) has a greater similarity to the PC learnt graph compared to the domain graph. 
Both AIT201 and AIT203, instead of P203 as in the PC graph, become confounders in this graph. 
With a root node as that in the domain graph, the CL learnt graph shows that FIT201 influences P201 and P203, instead of the AITs as in the domain graph. 
However, the CL learnt graph shows that the causal effects can be propagated from FIT201 to AIT201 via P203, and to AIT203 via P203 and AIT201, and to AIT202 via P203, AIT201 and AIT203. 
Importantly, the majority of the conditional independencies in the domain graph can be observed in each of the learned graphs. 
For instance, there is no direct edge between the pumps P201, P203 and P205. 
And MV201 remain isolated in the learnt graphs as well. 

\subsection{Causal Graphs in Stage-3: Ultrafiltration}
A total of nine DPs in Stage-3 can be found in historian data logs, three of them are continuous RVs. 
We have carefully investigated the distribution of the values of each of these RVs and convert them to suitable discrete RVs.  
Similar to other FITs, FIT301 can have two states, 0 to indicate no flow and 1 to indicate flow of water.
Similarly, LIT301 has three states; 0 1 and 2 to indicate Low, Medium and High level of water in tank T301. 
Also, we found that DPIT301 can have three states; 0, 1 and 2 to indicate Low, Medium and High level of differential pressure in the UV system. 

\subsubsection{Causal Domain Graph in Stage-3:} 
\begin{figure}[h]
     \centering
     \begin{subfigure}[b]{0.23\textwidth}
         \centering
         \includegraphics[width=\textwidth]{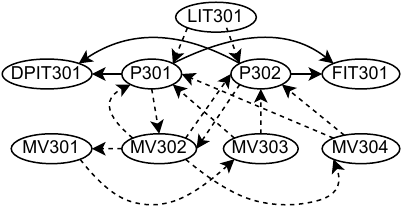}
         \caption{Causal domain graph}
         \label{fig:y equals x}
     \end{subfigure}
     \hfill
     \begin{subfigure}[b]{0.23\textwidth}
         \centering
         \includegraphics[width=\textwidth]{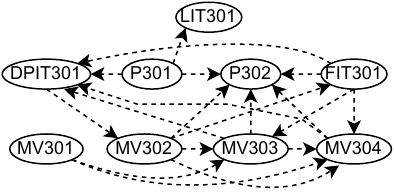}
         \caption{PC learnt graph}
         \label{fig:three sin x}
     \end{subfigure}
     \hfill
     \begin{subfigure}[b]{0.23\textwidth}
         \centering
         \includegraphics[width=\textwidth]{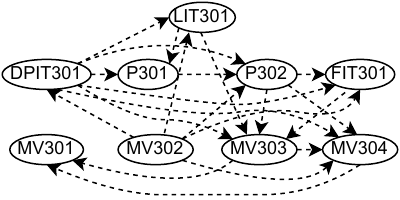}
         \caption{HC learnt graph}
         \label{fig:five over x}
     \end{subfigure}
     \hfill
     \begin{subfigure}[b]{0.23\textwidth}
         \centering
         \includegraphics[width=\textwidth]{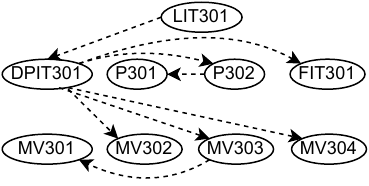}
         \caption{CL learnt graph}
         \label{fig:five over x}
     \end{subfigure}
        \caption{Causal graph in Stage-3 in SWaT. All the learnt graphs use BIC as scoring method.}
        \label{fig:swat-stage3-data-knowledge}
\end{figure}

Figure \ref{fig:swat-stage3-data-knowledge} shows the domain graph obtained in Stage-3. 
Based on domain knowledge, we find that P301 and P302 are physically coupled with DPIT301 and hence we have two edges denoting physical coupling, from P301 and P302 to DPIT301. 
FIT301 is physically coupled with the pipe through which P302 pushes water and hence we have indicated a physical coupling from P302 to FIT301. 
All other links in the domain graph are based on the control logic in Stage-3. 
For instance, the water level indicated by LIT301 controls both P301 and P302 to switch on or off. 
Indeed, activating these pump can depend on the other parameters present in the whole SWaT, here we focus only on Stage-3.  
Important to note here is that the control logic suggests that P301 can be activated provided MV302 is Open and MV302 is opened is considered as a preparation to switch on P301. 
Constructing a causal graph following the two principles may not allow us to create a DAG if all the dependencies be displayed individually. 
Further, if a casual graph is not DAG then parameter estimation is not possible. 
Therefore, though all possible edges are displayed, we shall drop a few from the domain graph by judiciously selecting them so that the impact is  minimal.  

\subsubsection{Parameter Estimation for the Causal Domain Graph in Stage-3:} 
\begin{figure}[h]
     \centering
     \begin{subfigure}[b]{0.23\textwidth}
         \centering
         \includegraphics[width=\textwidth]{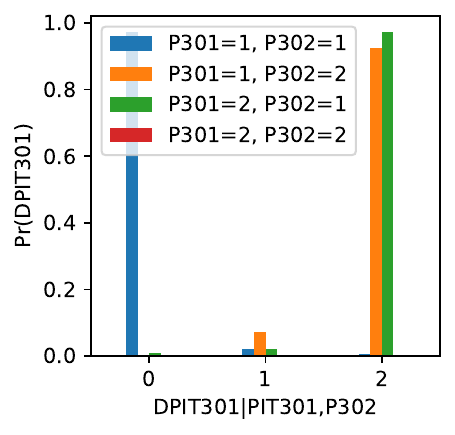}
         \caption{Parameter estimation for  $Pr(DPIT301|P301,P302)$}
         \label{fig:y equals x}
     \end{subfigure}
     \hfill
     \begin{subfigure}[b]{0.23\textwidth}
         \centering
         \includegraphics[width=\textwidth]{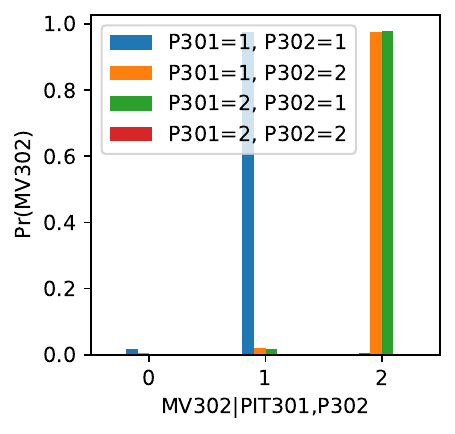}
         \caption{Parameter estimation for  $Pr(MV302|P301,P302)$}
         \label{fig:three sin x}
     \end{subfigure}
     \hfill
     \begin{subfigure}[b]{0.23\textwidth}
         \centering
         \includegraphics[width=\textwidth]{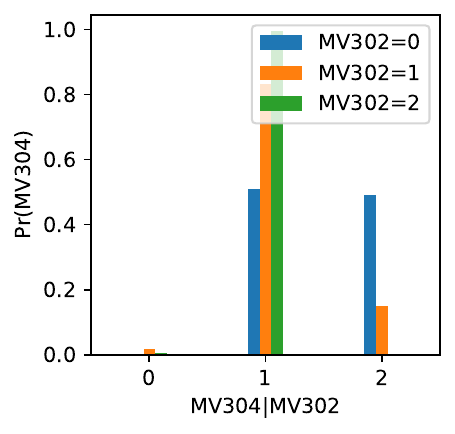}
         \caption{Parameter estimation for $Pr(MV304|MV302)$}
         \label{fig:five over x}
     \end{subfigure}
     \hfill
     \begin{subfigure}[b]{0.23\textwidth}
         \centering
         \includegraphics[width=\textwidth]{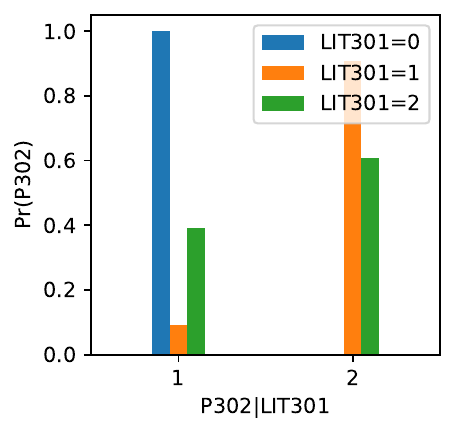}
         \caption{Parameter estimation for  $Pr(P302|LIT301)$}
         \label{fig:five over x}
     \end{subfigure}
        \caption{Parameter estimation in the causal domain graph in Stage-3 in SWaT.}
        \label{fig:swat-stage3-param-estimate}
\end{figure}

LIT301 is the only independent RV and the probabilities are $Pr(LIT301=0) \approx 0.02$, $Pr(LIT301=1) \approx 0.67$ and $Pr(LIT301=2) \approx 0.30$ as estimated by MLE.
When LIT301 is Low, none of P301 and P302 can be switched on and hence $Pr(P301 = 1 | LIT301 = 0) = 1.0$ and $Pr(P302 = 1 | LIT301 = 0) = 1.0$. 
Alternately, when LIT301 is not Low, P301 is switched on with probability 0.02, i.e., $Pr(P301 = 2|LIT301 = 1) \approx 0.02$, and it is switched off with probability about 0.97. 
Limiting the number of edges for analysis, we look into a RV that is affected by two other RVs. 
DPIT301 has two incoming edges, one each from P301 and P302 (Figure \ref{fig:swat-stage3-param-estimate}(a)). 
Therefore, a total of four possible combination of the states of the two pumps can be observed. 
The probability that $DPIT301 = 0$  is about 0.97 when none of the pumps are switched on. 
And, the probabilities that $DPIT301 = 2$, i.e., DPIT301 indicates High pressure, are about 0.92 and 0.97 when either P301 or P302 is On. 
Because the pumps cannot be switched on at the same time, the probability of DPIT301 in any of its state is undefined by MLE.

Similar to DPIT301, MV302 is also affected by the two pumps, P301 and P302 (Figure \ref{fig:swat-stage3-param-estimate}(b)). 
The probability that MV302 is Open given any one of the two pumps is switched on is about 0.97.
Also, it is Close with a probability of about 0.97 given both the pumps are switched off.
MV304 is affected by MV302 in turn (Figure \ref{fig:swat-stage3-param-estimate}(c)). 
Both the valves have three states.
So, for any given state of MV302, probability of MV304 is computed. 
The probability $Pr(MV304=1|MV302=2) \approx 0.99$, i.e., if MV302 is Open then MV304 is Close almost always. 
If MV302 is Close, then MV304 can be either Close or Open with almost equal probability (about 0.48). 
Looking at the edges associated with the pumps, P302 can be switched on with a higher probability (about 0.9) given LIT301 is Medium (Figure \ref{fig:swat-stage3-param-estimate}(d)).
It can also be switched on with a probability of about 0.6 even when LIT301 is High.
Thus, we looked into the parameters of four edges and the variation of the probability is observed which can validate the presence of each of the edges in the domain graph. 

\subsubsection{Causal Learnt Graphs in Stage-3:}
Figure \ref{fig:swat-stage3-data-knowledge}(b), Figure \ref{fig:swat-stage3-data-knowledge}(c) and Figure \ref{fig:swat-stage3-data-knowledge}(d) show the PC learnt, HC learnt and CL learnt graphs respectively. 
In general, each of the learnt graphs has a good number of both known and unknown edges compared to that in the domain graph. 
Opposing to domain knowledge, both PC and HC learnt graphs show that LIT301 is causally affected by P301, which is possibly because P301 in switch on state changes the reading of LIT301. 
Similarly, FIT301 is causally affecting P302 in the learned graph. 
Alternately, the impact of MV301 on MV303 can be seen in the learned graphs as that in the domain graph. 
A number of new causal relations can also bee seen in both the PC and HC learnt graphs.
For instance, FIT301 has a direct impact on MV304 in the PC graph. 
Opposing to the PC graph, HC graph shows that both MV303 and MV304 have causal impact on MV301. 

The CL learnt graph is showing that LIT301 has a direct impact on DPIT301. 
The graph indicates that the chain of causal impact can be used to derive that LIT301 has indirect impact on P301 and P302, as that in the domain graph.  
Further, DPIT301 has a direct impact on MV302, MV303 and MV304. 
None of the pumps has any impact on the valves. 
However, both the pumps and the valves are commonly affected by DPIT301. 
Such a causal relation may be a manifestation of  some external events that is not captured in the data set in Stage-3. 

\subsection{Causal Graphs in Stage-4: Dechlorination}
Stage-4 has nine RVs, where four (AIT401, AIT402, FIT401 and LIT401) are continuous RVs. 
Our analysis indicates that AIT401, AIT402, FIT401 and LIT401 can have two, three, two and three discrete values respectively after discretisation. 

\subsubsection{Stage-3 Causal Domain Graph:}
\begin{figure}[h]
     \centering
     \begin{subfigure}[b]{0.23\textwidth}
         \centering
         \includegraphics[width=\textwidth]{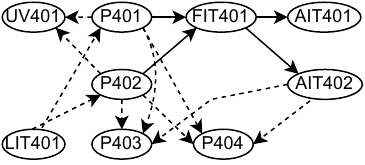}
         \caption{Causal domain graph.}
         \label{fig:y equals x}
     \end{subfigure}
     \hfill
     \begin{subfigure}[b]{0.23\textwidth}
         \centering
         \includegraphics[width=\textwidth]{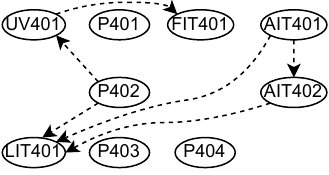}
         \caption{PC learnt graph.}
         \label{fig:three sin x}
     \end{subfigure}
     \hfill
     \begin{subfigure}[b]{0.23\textwidth}
         \centering
         \includegraphics[width=\textwidth]{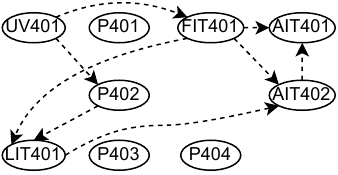}
         \caption{HC learnt graph.}
         \label{fig:five over x}
     \end{subfigure}
     \hfill
     \begin{subfigure}[b]{0.23\textwidth}
         \centering
         \includegraphics[width=\textwidth]{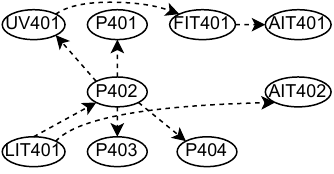}
         \caption{CL learnt graph.}
         \label{fig:five over x}
     \end{subfigure}
        \caption{Causal graphs in Stage-4 in SWaT. All the learnt graphs use BIC as scoring method.}
        \label{fig:swat-stage4-data-knowledge}
\end{figure}

Figure \ref{fig:swat-stage4-data-knowledge}(a) shows the causal domain graph in Stage-4. 
FIT401 is attached to the pipe that P401 or P402 pumps water through and the chemical sensors AIT401 and AIT402 are also attached to the same pipe.
Therefore, the edges from P401 or P402 to FIT401 and the edges from FIT401 to both AIT401 and AIT402 indicate physical coupling. 
The level of chemical indicated by AIT402 controls if P403 or P404 be switched on or not. 
So, these edges are due to control dependency. 
Like previous stages, LIT401 has control dependency on P401 and P402. 
According to control logic, chemical can be pumped using P403 or P404 into water only when water is pushed through the pipe using P401 or P402. 
Therefore, we have created control edges from P401 and P402 to both P403 and P404.  

\subsubsection{Parameter Estimation for Causal Domain Graph in Stage-4:}
\begin{figure}[h]
     \centering
     \begin{subfigure}[b]{0.23\textwidth}
         \centering
         \includegraphics[width=\textwidth]{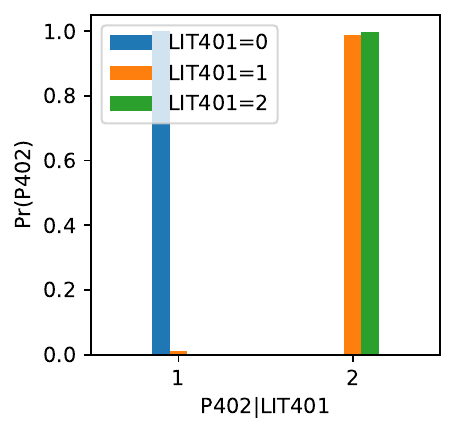}
         \caption{param estimate P402}
         \label{fig:y equals x}
     \end{subfigure}
     \hfill
     \begin{subfigure}[b]{0.23\textwidth}
         \centering
         \includegraphics[width=\textwidth]{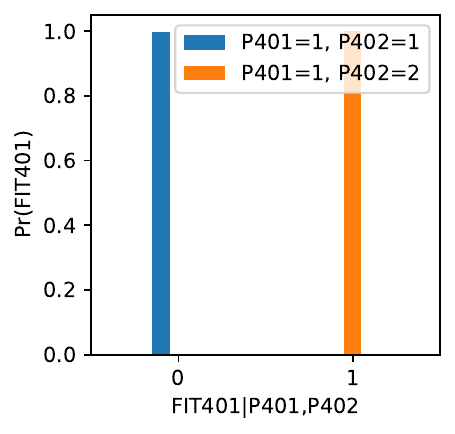}
         \caption{param estimate FIT401}
         \label{fig:three sin x}
     \end{subfigure}
     \hfill
     \begin{subfigure}[b]{0.23\textwidth}
         \centering
         \includegraphics[width=\textwidth]{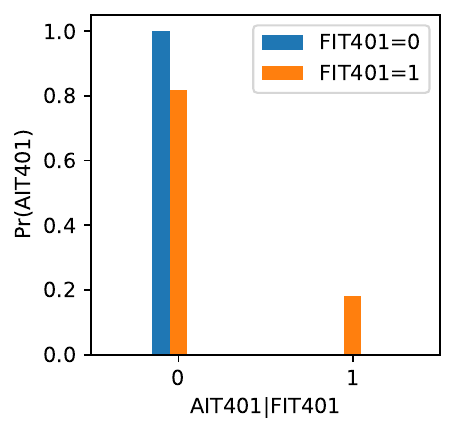}
         \caption{Param estimate AIT401}
         \label{fig:five over x}
     \end{subfigure}
     \hfill
     \begin{subfigure}[b]{0.23\textwidth}
         \centering
         \includegraphics[width=\textwidth]{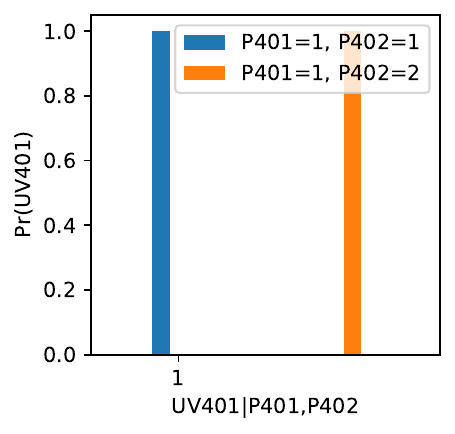}
         \caption{param estimate UV401}
         \label{fig:five over x}
     \end{subfigure}
        \caption{Parameter estimation in Stage-4 in SWaT.}
        \label{fig:swat-stage4-param-estimate}
\end{figure}

LIT401 having three discrete states is the independent RV in the domain graph in Stage-4. 
The parameters estimated using MLE for LIT401 are  $Pr(LIT401=0) \approx 0.02$, $Pr(LIT401=1) \approx 0.81$ and $Pr(LIT401=0) \approx 0.16$, which indicates that LIT401 being Medium is most probable. 
Similar to other Stages, the pumps P401 and P402 are affected by LIT401 (Figure \ref{fig:swat-stage4-param-estimate}(a)). 
The probability of P401 being Low given any state of LIT401 is 1.0, i.e., $Pr(P401=1|LIT401)=1.0$, which is due to the fact that P401 and P402 are back up for each other, and P401 is not seen to be activated during data collection. 

As expected, P402 is never switched on when LIT401 is indicating Low, i.e., $Pr(P402 = 1|LIT401=0) = 1.0$. 
When LIT401 is not Low, P402 is switched on with probability about 0.98.
FIT401 has joint causal impact of both P401 and P402 and therefore, four parameters, one for each combination of the states of the pumps, are estimated (Figure \ref{fig:swat-stage4-param-estimate}(b)). 
For instance, $Pr(FIT401=1|P401 = 1, P402 = 2) \approx 0.99$ states that if P402 is On and P401 is Off, then FIT401 indicates High flow of water.
AIT401 remains in state 0 as long as FIT401 is Low. 
The probability that AIT401 is in state 0 given FIT401=1 is about 0.81 and it is in state 1 otherwise (Figure \ref{fig:swat-stage4-param-estimate}(c)). 
Finally, UV401 is jointly affected by both P401 and P402 and hence four parameters are estimated (Figure \ref{fig:swat-stage4-param-estimate}(d)). 
But, because P401 is never switched on and two pumps cannot be switched on at the same time, there are only to parameters to be estimated. 
If none of the pumps are switched on, UV401 remain in state 0, i.e., $Pr(UV401 = 0 | P401 =1 , P402 = 1) = 1.0$. 
Alternately, if P402 is switched on, then the probability that UV401=1 is about 0.99, i.e., $Pr(UV401 = 1 | P401 = 1, P402 = 2) \approx 0.99$. 

\subsubsection{Stage-4 Causal Learnt Graphs:}
Figure \ref{fig:swat-stage4-data-knowledge}(b), Figure \ref{fig:swat-stage4-data-knowledge}(c) and Figure \ref{fig:swat-stage4-data-knowledge}(d) show the PC, HC and CL learnt graphs respectively in Stage-4. 
In general, each of the learnt graphs has a smaller number of edges compared to that in the domain  graph.  
Because P402 has been used, keeping P401 as backup, for the operations of the plant during the data collection process, both PC and HC learnt graphs keep P401 as isolated.
Both of these graphs show that P402 has a causal impact on LIT401. 
The edge between UV401 and P402 has different directions in PC and HC learnt graphs, i.e., even though the dependency relation is discovered, there are differences in the way one causes the other. 
The impact of UV401 on FIT401 is same in both the graphs.  
While UV401 is an independent RV in the HC graph, P402 is an independent RV in the PC graph. 
AIT401 is a confounder for LIT401 and AIT402 in the PC graph, whereas FIT401 is a confounder for AIT401 and AIT402 in the HC graph.
The pumps P403 and P404 (one is back up for the other) are isolated in both the PC graph and HC graph. 

In the CL learnt graph (Figure \ref{fig:swat-stage4-data-knowledge}(d)), we keep LIT401 as root node as that in the domain graph.
P402 has causal impact on four RVs, UV401, P401, P403 and P404 in this graph. 
UV401 in turn has impact on FIT401, and FIT401 further has impact on AIT401. 
Unlike the other two graphs, AIT401 and AIT402 has no relation between them in this graph, rather LIT401 has a causal impact on AIT402.

\subsection{Causal Graphs in Stage-5: Reverse Osmosis}
Stage 5 in SWaT has a total of thirteen RVs, where eleven are continuous RVs. 
Our analysis reveals that each of the chemical sensors, i.e., AITs, and the water level sensors, i.e., LITs, can have three discrete states; 0 -- Low, 1 -- Medium and 2 -- High. 
Each of the pressure sensors, i.e., PITs, can have two states; 0 -- Low, and 1 -- High.

\subsubsection{Causal Domain Graph in Stage-5:}
\begin{figure}[h]
     \centering
     \begin{subfigure}[b]{0.23\textwidth}
         \centering
         \includegraphics[width=\textwidth]{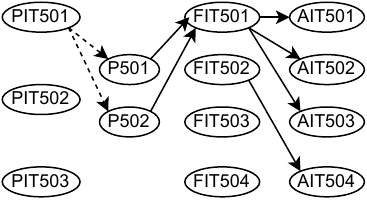}
         \caption{Causal domain graph}
         \label{fig:y equals x}
     \end{subfigure}
     \hfill
     \begin{subfigure}[b]{0.23\textwidth}
         \centering
         \includegraphics[width=\textwidth]{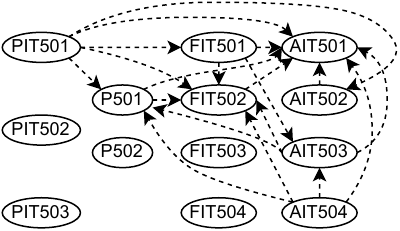}
         \caption{PC learnt graph.}
         \label{fig:three sin x}
     \end{subfigure}
     \hfill
     \begin{subfigure}[b]{0.23\textwidth}
         \centering
         \includegraphics[width=\textwidth]{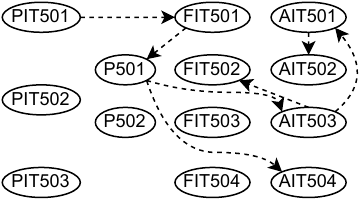}
         \caption{HC learnt graph.}
         \label{fig:five over x}
     \end{subfigure}
     \hfill
     \begin{subfigure}[b]{0.23\textwidth}
         \centering
         \includegraphics[width=\textwidth]{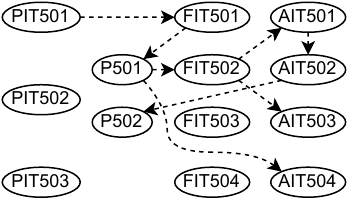}
         \caption{CL learnt graph.}
         \label{fig:five over x}
     \end{subfigure}
        \caption{Causal graphs in Stage-5 in SWaT. All the learnt graphs use BIC as scoring method.}
        \label{fig:swat-stage5-data-knowledge}
\end{figure}

The causal domain graph in Stage-5 (Figure \ref{fig:swat-stage5-data-knowledge}(a)) has relatively less number of control edges compared to the physical edges.  
Stage-5 has three PITs installed in different parts of the main water pipe, where PIT501 is at the first position in the pipe.
Hence, PIT501 is considered as an independent RV, which has control dependency on both the pumps P501 and P502 (one being backup for the other). 
Because both the flow meters, i.e., FITs, and the chemical sensors, i.e., AITs, are physically connected to the same pipe through which the pumps can push water, the edges from the pumps to the flow sensors and from the flow sensors to the chemical sensors are physical edges. 

\subsubsection{Parameter Estimation for the Causal Domain Graph in Stage-5:}
\begin{figure}[h]
     \centering
     \begin{subfigure}[b]{0.23\textwidth}
         \centering
         \includegraphics[width=\textwidth]{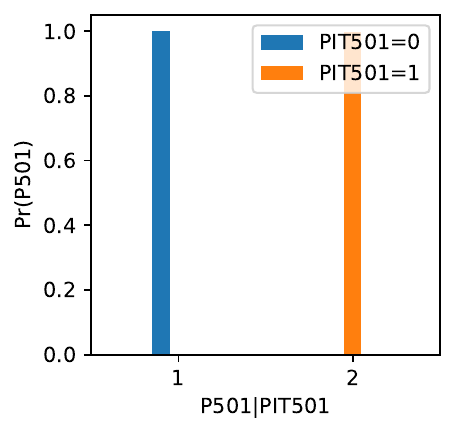}
         \caption{param estimate P501}
         \label{fig:y equals x}
     \end{subfigure}
     \hfill
     \begin{subfigure}[b]{0.23\textwidth}
         \centering
         \includegraphics[width=\textwidth]{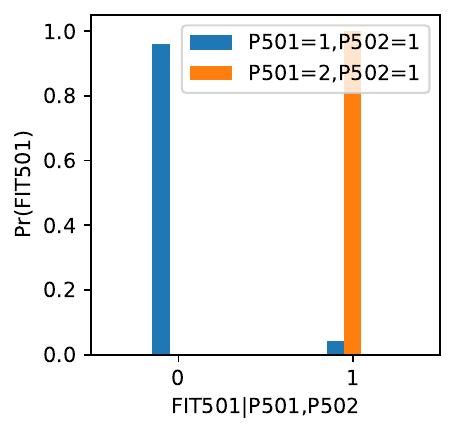}
         \caption{param estimate FIT501}
         \label{fig:three sin x}
     \end{subfigure}
     \hfill
     \begin{subfigure}[b]{0.23\textwidth}
         \centering
         \includegraphics[width=\textwidth]{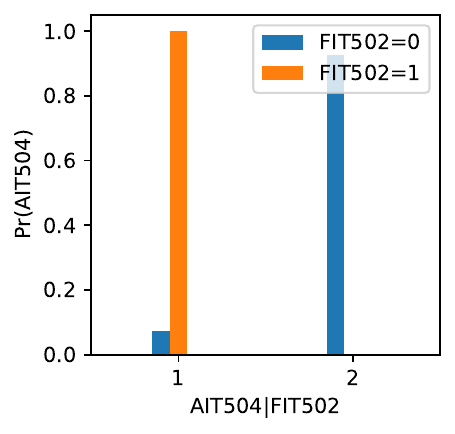}
         \caption{Param estimate AIT504}
         \label{fig:five over x}
     \end{subfigure}
     \hfill
     \begin{subfigure}[b]{0.23\textwidth}
         \centering
         \includegraphics[width=\textwidth]{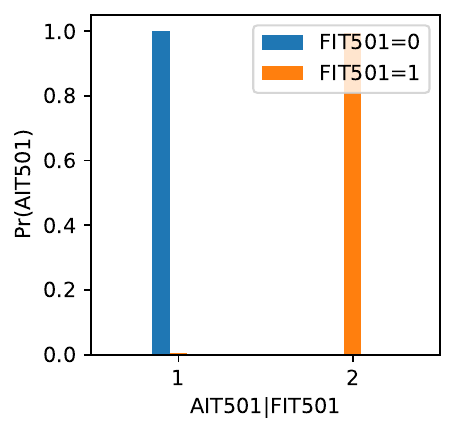}
         \caption{param estimate AIT501}
         \label{fig:five over x}
     \end{subfigure}
        \caption{Parameter estimation in Stage-5 in SWaT.}
        \label{fig:swat-stage5-param-estimate}
\end{figure}

The simple probability that PIT501 (an independent RV in Stage-5) being Low is about 0.03 and it is being High is about 0.96.
The probability $Pr(P501=2|PIT501=0) = 0$ indicates that P501 cannot be switched on if PIT501 indicates Low pressure (Figure \ref{fig:swat-stage5-param-estimate}(a)).
Alternately, given PIT501 is High, the pump P501 can be switched on with a probability of about 0.99. 
P502 being a backup pump is not switched on irrespective of pressure level indicated by PIT501.
The probability 0f FIT501 indicating Low is 0 if P502 is switched on, i.e., $Pr(FIT501 = 0|P501 = 2, P502=1) = 0$ (Figure \ref{fig:swat-stage5-param-estimate}(b)).  
Alternately, given that both the pumps are switched off, FIT501 indicates High is about 0.04, i.e., $Pr(FIT501 = 1|P501=1, P502 = 1) \approx 0.04$. 

For FIT502 (another independent RV), $Pr(FIT502=0) \approx 0.03$ and $Pr(FIT502=0) \approx 0.96$. 
Given that FIT502 is Low, AIT504 is always 0, i.e., $Pr(AIT504=0|FIT502 = 0) = 1.0$ (Figure \ref{fig:swat-stage5-param-estimate}(c)). 
The probability that AIT501 is in state 1 given FIT501 indicates High is about 0.99, i.e., $Pr(AIT501 = 1|FIT501 = 1) \approx 0.99$, and it remains in state 0 if there is no flow (Figure \ref{fig:swat-stage5-param-estimate}(d)). 
A similar set of parameters are seen in case of AIT502.
However, for AIT503, the probability that it is in state 1 is about 0.99 irrespective of flow level indicated by FIT501.  

\subsubsection{Causal Learnt Graphs in Stage-5:}
Figure \ref{fig:swat-stage5-data-knowledge}(b), Figure \ref{fig:swat-stage5-data-knowledge}(c) and Figure \ref{fig:swat-stage5-data-knowledge}(d) show the PC, HC and CL learnt graphs respectively in Stage-5. 
PIT501 has been an independent RV in both the PC and HC graphs, as that in the domain graph. 
While PIT501 can be seen to have causal impact on five RVs in the PC graph, it has impact only on FIT501 in the HC graph. 
AIT501 is impacted by seven RVs in the PC graph and it does not have any impact on other RVs. 
The causal dependencies in the PC graph is relatively complex compared to that in the HC graph.
Unlike in the PC graph, a variable is impacted only by one variable and a maximum of two variables are impacted by a single RV in the HC graph. 

The CL learnt graph is relatively more similar to the HC learnt graph. 
FIT501 has its causal impact only on P501, and P501 in turn impacts two other RVs. 
Two chemical sensors AIT501 and AIT503 are causally affected by FIT502. 
The edge from AIT502 to P502 has not been seen in any of the PC or HC graphs. 
Finally, PIT502, PIT503 and FIT504 remain isolated in all the four causal graphs in this stage. 

\subsection{Causal Graphs in Stage--6: Backwash}
\begin{figure}[h]
     \centering
     \begin{subfigure}[b]{0.2\textwidth}
         \centering
         \includegraphics[width=\textwidth]{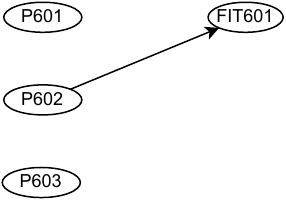}
         \caption{Causal domain graph.}
         \label{fig:y equals x}
     \end{subfigure}
     \hfill
     \begin{subfigure}[b]{0.2\textwidth}
         \centering
         \includegraphics[width=\textwidth]{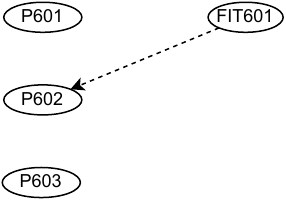}
         \caption{PC learnt graph.}
         \label{fig:three sin x}
     \end{subfigure}
     \hfill
     \begin{subfigure}[b]{0.2\textwidth}
         \centering
         \includegraphics[width=\textwidth]{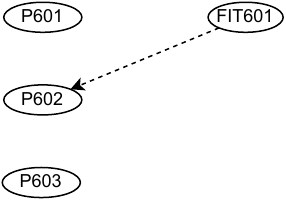}
         \caption{HC learnt graph.}
         \label{fig:five over x}
     \end{subfigure}
     \hfill
     \begin{subfigure}[b]{0.2\textwidth}
         \centering
         \includegraphics[width=\textwidth]{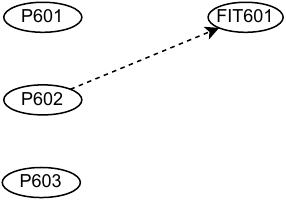}
         \caption{CL learnt graph.}
         \label{fig:five over x}
     \end{subfigure}
        \caption{Causal graphs in Stage-6 in SWaT. All the learnt graphs use BIC as scoring method.}
        \label{fig:swat-stage6-data-knowledge}
\end{figure}

Four RVs are used to collect historian data logs, where one is a continuous RV. 
Similar to other stages, our analysis reveals that FIT601 can have two states. 
The causal domain graph in Stage-6 (\ref{fig:swat-stage6-data-knowledge}(a)) has only one edge, i.e., (P602, FIT601), which is a physical edge. 
Estimating the parameters, P602 can be switched on and off with the probabilities about 0.01 and 0.99 respectively (plots not reported due to space constraint). 
The probability that FIT601 indicates Low given that P602 is Off is about 0.99. i.e., $Pr(FIT601 =0 | P602 = 1) = 0.99$. 
And, it can show High given P602 is On is about 0.76. Figure \ref{fig:swat-stage6-data-knowledge}(b),  Figure \ref{fig:swat-stage6-data-knowledge}(c) and Figure \ref{fig:swat-stage6-data-knowledge}(d) show 
the PC, HC and CL learnt graphs respectively. 
In the PC and HC graph, the edge between P602 and FIT601 is reversed compared to the domain graph. 
Because P602 is considered as root, the CL graph has the edge as in the domain graph. 

\subsection{Summary of the Dynamics in the Learnt Graphs in the Six Stages}
In summary, the causal learn graphs have non-negligible abilities to represent the relationships among the DPs in each of the six stages in SWaT. 
Because of two types of edges (control and physical) in the causal domain graphs, we can distinguish the stages of SWaT based on the abilities to discover the causal relations learnt by the algorithms.  
Stage-1 and Stage-6 are primarily controls water storage and flow direction and majority of the learnt graphs can replicate the physical edge. 
Relatively more important (or not-seen-before) causal relations can be observed in the other stages where chemical injection or removal is performed. 

For instance, in Stage-2, both PC and HC learnt graphs reveal that one chemical sensor has impact on other chemical sensor. 
In Stage-3, while the PC learn graph reveals differential pressure sensor is affected by several other RVs including pumps, flow meters and valves, the HC learnt graph shows that almost same set of DPs are connected to the DPIT (though directions are differing). 
Thus, we find that each of the stages can exhibit a different set of causal relations based on their functionalities in the plant like RO, Dechlorination, Ultrafiltration and Chemical Dosing. 

\section{Discovering The Impacted {DPs} using Causal Graph}
\label{sec:attack-space}
In this section, we shall explore one of the possible usecases of the learnt causal graphs in a CPS.
Our usecase in this paper is to discover the Impacted DPs $\mathcal{A}_{d|s}$ given a cyber attack that is launched on Target DPs $\mathcal{A}_{s}$.
For a better comprehensibility, we do not alter the names of the RVs while exploring the impact. 
We shall make use of a set of well known nine attacks \cite{sridharAsiaCCS2016, Yoong2021CyberSecurity} in SWaT for the first time in this paper. 

\subsection{Attacks on SWaT}
In this section, we shall analyze and derive the Targeted and the Impacted DPs of the attacks in \cite{sridharAsiaCCS2016, Yoong2021CyberSecurity}. 
Let us consider four classes of attacks in a CPS based on Targeted and Impacted DPs: 

\begin{itemize}
    \item Target:Single, Impact:Single (TSIS): $\mathcal{A}_s$ contains the DPs that are belonging to one Stage only, say $S_i$, and $\mathcal{A}_{d|s}$ contains the DPs that are belonging to one stage only, say $S_j$, where $S_i$ may be same as $S_j$. 
    
    \item Target:Single, Impact:Multiple (TSIM): 
    $\mathcal{A}_s$ contains the DPs that are belonging to one Stage only, say $S_i$, and $\mathcal{A}_{d|s}$ contains the DPs that are belonging to more than one stage, say $\mathbb{S}_j$, where $S_i$ may belong to $\mathbb{S}_j$. 
    
    \item Target:Multiple, Impact:Single (TMIS): $\mathcal{A}_s$ contains the DPs that are belonging to more than one Stage, say $\mathbb{S}_i$, and $\mathcal{A}_{d|s}$ contains the DPs that are belonging to one stage only, say $S_j$, where $\mathbb{S}_i$ may contain $S_j$.
    
    \item Target:Multiple, Impact:Multiple (TMIM): Both $\mathcal{A}_s$ and $\mathcal{A}_{d|s}$ contains the DPs that are belonging to more than one stage, say $\mathbb{S}_i$, and $\mathbb{S}_j$ respectively, where $\mathbb{S}_i \cap \mathbb{S}_j$ need not be empty.
\end{itemize}

We consider two attacks, denoted by ($\mathcal{A}_s^1$, $\mathcal{A}_{d|s}^1$) and ($\mathcal{A}_s^2$, $\mathcal{A}_{d|s}^2$), are different if either $(\mathcal{A}_s^1 \cup \mathcal{A}_s^2) - (\mathcal{A}_s^1 \cap \mathcal{A}_s^2) \neq \phi$, or $(\mathcal{A}_{d|s}^1 \cup \mathcal{A}_{d|s}^2) - (\mathcal{A}_{d|s}^1 \cap \mathcal{A}_{d|s}^2) \neq \phi$ or both holds. 
Such an assumption is inline with the fact that a particular set of Targeted DPs can be attacked in a way that the Impacted DPs are different. 
For instance, if both P101 and P102 switched off then the Impacted DPs can be different from those if both of them are switched on together due to an attack.

Table \ref{tab:attack-target-intent-using-causal} shows all nine attacks (used in \cite{sridharAsiaCCS2016, Yoong2021CyberSecurity}) that we use in this paper. 
We consider these attacks in light of the Targeted DPs, and the Impacted DPs.  
Table \ref{tab:attack-target-intent-using-causal} reports the Impacted DPs as can be derived from the existing works in \cite{sridharAsiaCCS2016, John2018ACSAC} for the first time in this paper along with the ones discovered by our proposed system. 


In Attack 1, MV101 is attacked where its state is changed frequently when LIT101 is High. 
The aim is to see if LIT101 responds appropriately or if MV101 allow such sudden change in its state. 
Thus, $\mathcal{A}_s = \{MV101\}$ and $\mathcal{A}_{d|s} = \phi$
and this attack belongs to TSIS category.  
In Attack 2, the state of P101 is changed given LI101 and MV201 are in some particular states (e.g., Medium and Open resp.). 
For instance, if LIT101 is already Low or Medium, then switching on P101 can cause tank T101 to underflow. 
Thus, $\mathcal{A}_s = \{P101\}$ and $\mathcal{A}_{d|s} = \{LIT101\}$ and this attack belongs to TSIS category.

In Attack 3, P101 and P102 are the Target DPs, where the states of these DPs are changed given certain states MV201 and LIT301 (e.g., Close and High resp.). 
The attack aims to check the resilience of the water pipe in Stage-2, i.e., there is no particular DP is expected to be impacted. 
Thus, $\mathcal{A}_s = \{P101, P102\}$ and $\mathcal{A}_{d|s} = \phi$ and this attack belongs to TSIS category.
In Attack 4, three DPs (MV201, P101 and P102) belong to Targeted DPs.
Note that the DPs in Target DPs belong to two stages, Stage-1 and Stage-2. 
The attack can be launched when LIT301 is in a particular state, e.g., Medium. 
The attack does not aim to impact on any other DPs than the target DPs.
Thus, $\mathcal{A}_s = \{MV201, P101, P102\}$ and $\mathcal{A}_{d|s} = \phi$ and this attack belongs to TMIM category.
In Attack 5, LIT301 is in Target DPs, where its state is changed from Low to High or from High to Low suddenly. 
The impact of this attack is expected to be observed in both MV201 and P101.
Thus, $\mathcal{A}_s = \{LIT301\}$ and $\mathcal{A}_{d|s} = \{MV201, P101\}$ and this attack belong to TSIM category.  

\begin{table}[h]
  \begin{tabular}{ | p{0.2in} | p{1.3in} | p{0.4in} |  p{1.2in} | p{1.0in} | p{0.9in} |}
    \hline
    Sl. NO. & $\mathcal{A}_s$  & $\mathcal{A}_{d|s}$ as in \cite{sridharAsiaCCS2016} & $\mathcal{A}_{d|s}$ as in \cite{John2018ACSAC} & $\mathcal{A}_{d|s}$ as in domain knowledge graph & $\mathcal{A}_{d|s}$ as in learned causal graph \\ \hline
    
    1 & Change state of MV101, given a state of LIT101 & None & MV201, \{P201, ..., P208\}, \{P601, ..., P603\} &  FIT101 & FIT101  \\ \hline
    
    2 & Change states of P101, given a state pair of LIT101 and MV201 & LIT101 &  MV201, \{P201, ..., P208\}, \{P601, ..., P603\} & $*$ & MV201 \\ \hline
    
    3 & Change states of both P101 and P102, given a state pair of MV201 and LIT301 & Water pipe in stage 2 & MV201, \{P201, ..., P208\}, \{P601, ..., P603\} & $*$ & MV201 \\ \hline
    
    4 & Change states of MV201, P101 and P102, given a state of LIT301 & None & MV201, \{P201, ..., P208\}, \{P601, ..., P603\} & $*$ & MV201  \\ \hline 
    
    5 & Change state of LIT301 & MV201, P101 &  --  & P301 and P302 & -- \\  \hline
    
    6 & Change state 0f P203, given a stat pair of MV201 and AIT202 & None & MV101, P101, P102, MV201, \{P201, P202, P205, ..., P208\}, \{MV301, ..., MV304\}, P301, P302 & $*$ & AIT201, FIT201 \\ \hline 
    
    7 & Change state of AIT202, given a state of MV201 & P203 & MV201, \{P201, P202, P205, ..., P208\} & P203 & -- \\ \hline
    
    8 & Change state of P402, given a state pair of P401 and P501 & Water pipe in state 4 & \{MV301, ..., MV304\}  & FIT401, UV401, P403 and P404 & UV401 \\ \hline
    
    9 & Change state of FIT501, given a state of P401 & None & -- & AIT501, AIT502 and AIT503 & FIT401, AIT501, PIT501 \\ \hline
  \end{tabular}
   \caption{Attack target and attack impact  identification.}
  \label{tab:attack-target-intent-using-causal}
\end{table}

In Attack 6, only P203 is in Targeted DPs, where its state is changed given particular states of MV201 and AIT202, i.e., all these DPs are in Stage-2. 
This attack is not expected to have impact on any other DPs and hence it belongs to TSIS category. 
In Attack 7, only AIT202 is attacked, where the state of AIT202 is changed given a particular state of MV201. The attack is expected show its impact on P203 and hence it belongs to TSIS category. 
In Attack 8, P402 belong to Target DPs, where its state is changed given particular states P401 and P501. 
The attack is aimed is to check resilience of the pipe segment in Stage-4.
Thus, this attack belong to TSIS category. 
In Attack 9, FIT501 belongs to Target DPs, where its value is changed when P401 is in a particular state, e.g., On. 
This attack is not expected to show its impact on any other DPs, so it belongs to TSIS category. 

An important and interesting factor that is common all these nine attacks is the preconditions.  
It is therefore important to investigate the probability of such preconditions using inference on the causal learnt graphs.

\subsection{Causal Graph to Discover Attack Target and Attack Impact}
In this section, we shall make use of inference on causal graphs to identify the Impacted DPs due to an attack. 

To identify Impacted DPs, i.e., $\mathcal{A}_{d|s}$, we consider a simple algorithm. 
For each 1-hop neighbour $v_j$ of $v_i\in \mathcal{A}_s$, use inference on a causal learnt graph to estimate the probability of a particular state of $v_i$ given each state $s_l$ of $v_j$. 
If there exist a state $s_k$ of $v_i$ such that $Pr(v_i = s_k | v_j = s_l)>= \theta$, where $\theta$ is a threshold, then $v_j$ belongs to $\mathcal{A}_{d|s}$. 
Let us consider a higher value for the threshold, say $\theta =0.9$; a higher threshold indicates a higher level of confidence of impact.  

\subsubsection{Attack on MV101} 
In Attack 1, both the target and the precondition involves the DPs only in Stage-1.
Therefore, we consider the causal learnt graph in Stage-1. 
Both P101 and FIT101 are 1-hop neighbour of MV101 as in the PC learnt graph (Figure \ref{fig:swat-stage1-data-knowledge}(b)), i.e., the graph learnt using PC algorithm. 
Given FIT101=0, the probability of MV101=1 is about 0.98 and hence FIT101 is in $\mathcal{A}_{d|s}^1$. 
Given P101 = 1, the probability of MV101=1 is about 0.49 and given P101 = 2, the probability of MV101=2 is about 0.80. 
Hence, we consider P101 as not in $\mathcal{A}_{d|s}^1$. 
The result is shown in Table \ref{tab:attack-target-intent-using-causal}. 
Additionally, Table \ref{tab:attack-target-intent-using-causal} shows the Impacted DPs derived from the causal domain graph. 

\subsubsection{Attack on P101} 
\begin{figure}[h]
     \centering
     \begin{subfigure}[b]{0.7\textwidth}
         \centering
         \includegraphics[width=\textwidth]{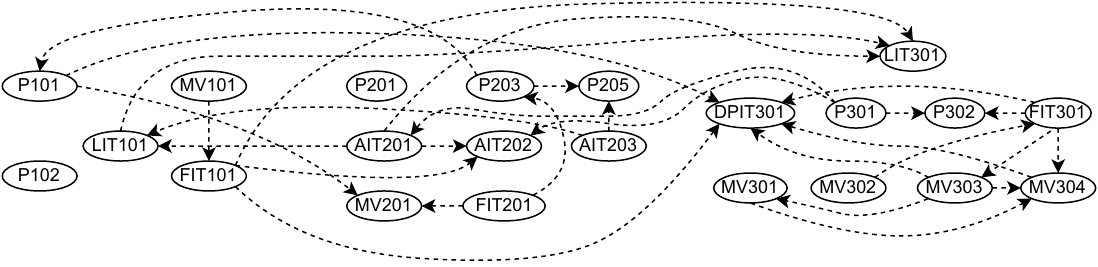}
     \end{subfigure}
     \caption{Causal learnt graph using DPs in Stage-1, Stage-2 and Stage-3 using PC algorithm (i.e., cs).}
     \label{fig:swat-stage123-causal-learned-cs}
\end{figure}

In Attack 2, the DPs in $\mathcal{A}_s^2$ and the precondition involves P101, LIT301 and MV201, i.e., it involves Stage-1, Stage-2 and Stage-3. 
Therefore, we first generate the causal learnt graph (Figure \ref{fig:swat-stage123-causal-learned-cs}) and then apply causal inference to identify the Impacted DPs. 
The PC graph using the DPs in three stages is significantly different from the graphs created using the DPs in each of the individual stages of Stage-1, Stage-2 and Stage-3. 
For instance, 
P203 has 1-hop causal impact on AIT201, FIT201, AIT203 in the causal graph learned using the DPs only in Stage-2 (Figure \ref{fig:swat-stage2-data-knowledge}(b)), whereas it has causal impact on P101 and P205. 

Applying inference on the PC learnt graph in Figure \ref{fig:swat-stage123-causal-learned-cs}, we can find that P101 (the Target DP in this case) has two direct neighbours, MV201 and DPI301, where $Pr(P101=1|MV201 = 1) = 0.99$ and there is no state of P101 for which the probability is more than $\theta$ given any state of DPIT301. 
Hence, only MV201 is in $\mathcal{A}_{d|s}^2$.  


\subsubsection{Attack on P101 and P102}
In Attack 3, $\mathcal{A}_{s}^3 = \{P101, P102\}$. 
Like Attack 2, there is no DP in $\mathcal{A}_{d|s}^3$ according to the causal domain graph in Stage-1. 
The DPs in $\mathcal{A}_{s}^3$ and the precondition are same as that in case of Attack 2. 
We apply same inference on the PC learnt graph shown in Figure \ref{fig:swat-stage123-causal-learned-cs}. 
P102 is an isolated DP in the causal graph and hence it is not used in the inference. 
Thus, $\mathcal{A}_{d|s}^3 = \{MV201\}$, as in case of Attack 2. 

\subsubsection{Attack on P101, P102 and MV201}
In attack 4, $\mathcal{A}_{s}^4 = \{P101, P102, MV201\}$ and it has precondition on LIT301, and hence we use the PC learnt graph as in case of Attack 2 and Attack 3 (Figure \ref{fig:swat-stage123-causal-learned-cs}). 
Because MV201 has no out going edge and P102 is an isolated node, the attack impact is only based on the outgoing edges of P101.
Hence, $\mathcal{A}_{d|s}^4 = \{MV201\}$.

\subsubsection{Attack on LIT301}
In Attack 5, LIT301 is in $\mathcal{A}_s$ without having any precondition and therefore, we consider the PC graph in Stage-3 (Figure \ref{fig:swat-stage3-data-knowledge}(b)).  
LIT301 has no outgoing edge in this causal graph, whereas it has only one incoming edge from P301.
So, $\mathcal{A}_{d|LIT301} = \phi$.  

\subsubsection{Attack on P203} 
In Attack 6, $\mathcal{A}_s^6 = \{P203\}$ having precondition on both MV201 and AIT202 and hence we consider the PC learnt graph in Stage-2 (Figure \ref{fig:swat-stage2-data-knowledge}(b)). 
Note that the graph has a cycle and we have removed a minimum number of edge selected judiciously from the graph in order to apply causal inference. 
The edge (AIT202, AIT201) is removed.
In the resulting PC learnt graph, P203 has three outgoing edges to AIT201, AIT203 and FIT201. 
Applying inference, we find that $Pr(P203 = 2 | AIT201 = 2) \approx 0.98$ and $Pr(P203 = 2 | FIT201=1) \approx 0.99$. 
The probability of any state of P203 given any state of AIT203 is less than the threshold $\theta = 0.9$. 
Thus, $\mathcal{A}_{d|s}^6 = \{AIT201, FIT201\}$.

\subsubsection{Attack on AIT202} 
In Attack 7, $\mathcal{A}_s^7 =\{AIT202\}$ and the precondition is on MV201.
So, we consider the PC learnt graph in Figure \ref{fig:swat-stage2-data-knowledge}(b). 
As in Attack 6, the edge (AIT202, AIT201) is removed from the graph. 
Hence, $\mathcal{A}_{d|s}^7 =\phi$.

\subsubsection{Attack on P402}
\begin{figure}[h]
     \centering
     \begin{subfigure}[b]{0.8\textwidth}
         \centering
         \includegraphics[width=0.9\textwidth]{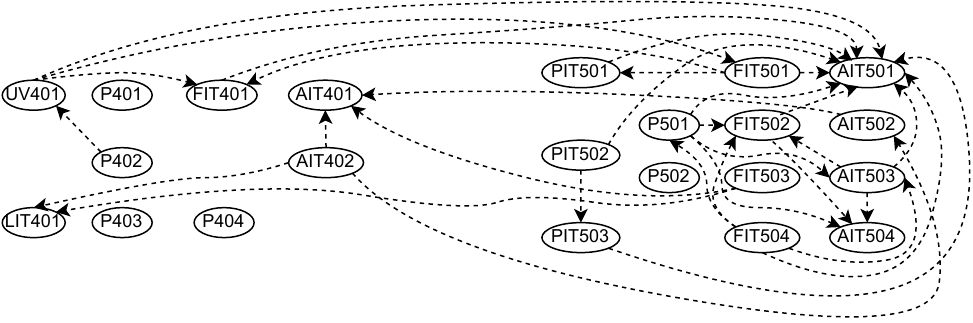}
         
     \end{subfigure}
     \caption{Learned causal graph using DPs in stage 4 5 using cs.}
     \label{fig:swat-stage45-causal-learned-cs}
\end{figure}

In Attack 8, $\mathcal{A}_s = \{P402\}$ and the precondition is on both P401 and P501. 
Therefore, we generate the PC learn graph (shown in Figure \ref{fig:swat-stage45-causal-learned-cs}) considering the DPs in Stage-4 and Stage-5 (i.e., a total of 22 DPs). 
Because P402 has only one out going edge to UV401, we apply inference on this node and it turns out that $Pr(P402=1|UV401 = 0) \approx 0.93$.
Hence, $\mathcal{A}_{d|s}^8 = \{UV401\}$. 

\subsubsection{Attack on FIT501}
In Attack 9, $\mathcal{A}_s = \{FIT501\}$ and the precondition is on P401.
So, we consider the PC learnt graph in Figure \ref{fig:swat-stage45-causal-learned-cs}. 
FIT501 has three outgoing edges to FIT401, PIT501 and AIT501. 
Applying inference on FIT501 given AIT501, we find that $Pr(FIT501=1|AIT501 = 2) \approx 0.98$. 
Similarly, $Pr(FIT501 = 0| PIT501=0) \approx 0.93$ and $Pr(FIT501 = 1|FIT401 = 1) \approx 0.99$. 
Therefore, $\mathcal{A}_{d|s}^9 = \{FIT401, PIT501, AIT501\}$. 

\subsection{Discussion}
\label{subsec:discussion}
Our proposed system, called ICCPS, makes use of either domain knowledge or historian data logs to build the causal graph consisting of the DPs in the system. 
While the principles of control dependency and physical coupling is easy to represent in a domain knowledge based causal graph, i.e., the domain graphs, it requires a significant manual effort to reason about every edge that can be introduced into the graph. 
Nevertheless, such a graph can become well suited in the scenarios where the historian data logs are not available. 
For instance, the domain graphs can be useful in order to follow Security-by-Design principle in a CPS, where the actual CPS is not yet operational or with legacy systems where only certain controllers, like SCADA, is operational and there is no provision of collecting historian data logs. 

One the other hand, the causal graphs learned using historian data logs can be useful in the modern CPSs where a separate historian can be expected to collect sensor readings or actuator states at a regular interval and the historian data logs are available to trusted third parties for advanced analytics, like the one proposed in this paper. 
In this case, ICCPS expects that the data set uniquely identifies each of the DPs in a CPS.  
ICCPS does not require any description for any of its DPs in the CPS. 
However, it requires all the DPs to represent discrete RVs. 
Therefore, if there is any continuous RV then we need to discretise the RV before applying it to the ICCPS. 
Essentially, the performance of ICCPS can vary depending on the discretisation of such RVs.

Our proposed ICCPS system can be applied similar other domains, like water distribution systems and weather monitoring systems, with no or very limited effort. 
The system would require certain knowledge of discretising each of the continuous RVs and this can be considered as the only limitation of applying ICCPS in a CPS system. 
We believe that ICCPS can be extended to perform other advanced analytics, like developing security test cases and discovering the DPs that have a higher potential to get affected as a result of a cyber attack on a CPS so that an efficient recovery strategy can be build up.


\section{Related Works} 
\label{sec:rel-works}

Among other, we consider the studies that are related to the generation of attacks, the detection of attack on a particular set of DPs, the identification of the DPs are attacked and the prevention of attacks on CPSs. 
First step in cyber attack is cyber reconnaissance which can be achieved by analysing Internet based intelligence or collecting some sample operational data logs or direct interaction with the target system \cite{mazurczyk2021cyber}. 
On the other hand, understanding the intention of an attacker can be achieved by developing a honeypot involving PLCs \cite{lopez2020honeyplc}. 

No matter how an attack is planned, actual attack generation and its detection has been one of the major focus in the research community. 
Generating appropriate false data and its injection into electric smart meters in a CPS like power grid \cite{liu2011false}, or injecting it into sensors or actuators to alter an expected state of a CPS \cite{cardenas2011attacks}, or attacking the control logic in PLCs in CPSs like SWaT \cite{adepu2016investigation} have been studied with an aim of creating security test cases, for example. 
Recently, automatic fuzzing is also proposed to generate a large number of security test cases \cite{chen2019learning}. 

Identification of the DPs under attack in an operational CPS can be achieved by constructing a timed automata \cite{lin2018tabor} or by analysing attack tree \cite{falco2018master, barrere2020fault}. 
Also, the deviation in the readings of a sensor, for instance, while modeling it using autoregression can also be used to detect stealthy attacks \cite{urbina2016limiting}. 
Further, creating a model of CPS with an aim of identifying the DPs with a higher risk \cite{hau2020evaluating} or analysing network traces to identify the packets carrying malicious or false data \cite{JohnRAID2021} have been proposed. 
Detection of anomalous state of a CPS, rather than a set of DPs in a CPS, has been proposed by techniques like implementing system \emph{invariants} \cite{adepu2016argus, adepu2018distributed} and by leveraging functional and non-functional requirements of a CPS to create a set of rules during the design phase of a system \cite{khan2017armet, yilmaz2018attack, das2020anomaly}. 

Finally, we discuss a number of studies that aims to predict a set of DPs that can be the direct victims of an attack in a CPS. 
We shall focus on those studies that make use of some kind of graph corresponding to a CPS. 
Construction and analysis of attack graphs have been an attractive research area, where the focus has been in the areas like the verification of attack graph in simulated environment \cite{hill2017verifying}, the augmentation of traditional attack graphs (called hybrid attack graphs) with additional node properties and edge weights \cite{nichols2017introducing}, the verification of security policies using attack graphs \cite{west2019critical}, or the assignment of a criticality score to the nodes or the edges in the graph \cite{kavallieratos2020attack, abdallah2020behavioral}.  

Various types of analytics have been proposed on the attack graphs, starting from identifying critical attack paths \cite{spathoulas2021attack}, to predict the next probable DP as a result of compromising a DP \cite{hoff2021creating}, to identify easy targets yet achieve maximum damage \cite{few2021case}. 
Further, a CPS is can be modeled in a multi-layered graph in order to identify the components that can be compromised in a coordinated manner to maximize the damage or introduce significant delay in a mission critical system \cite{pelissero2021model, barrere2021analysing}. 
A number of tools has been developed recently to automate certain tasks involving attack graphs, ranging from \cite{li2022strategies} enabling parallel programming and applying centrality measures in large attack graphs, to \cite{buczkowski2022optimal} for analysing probabilistic attack graphs, to \cite{al2019a2g2v, li2021system} for automatic graph generation, to \cite{ibrahim2019automatic} for automatic generation of hybrid attack graph. 

The work that is most relevant to the one presented in this paper is in \cite{John2018ACSAC}. 
This work has proposed to consider sensors as entities to convert physical environment to cyber information and actuators as entities to convert cyber information into physical environment. 
Modeling a CPS in terms of information flow between cyber and physical component allows to find a dependency among the components, which can then be represented in a data flow graph. 
Essentially, construction of such a graph requires to analyse control logic executed in a set of PLCs.  
An attacker in this design considers only sensors (represented as a node in the graph) as attack points, (though unrealistic, yet we assume communication links are secure for simplicity). 
Thus, a data flow graph can help to identify potential impacted components the CPS as result of injecting false data into sensor readings. 
In this paper, we consider a fundamentally different approach compared to that in \cite{John2018ACSAC}, where control dependency and physical coupling is converted to dependency relationship among the DPs in causal graph while constructing the domain graphs.
And, consider historian data logs while learning the causal graphs. 
Finally, we apply causal inference on the causal graphs for discovering the  impacted components (i.e, DPs).

\section{Conclusion}
\label{sec:conclusion}
In this paper, we have aimed to address the problem of identifying impact points of a (cyber) attack that can be launched on certain target point(s) in a Cyber Physical System (CPS) with the help of causal graphs. 
We have adopted a set of basics to construct the causal graphs in the domain of CPS that has both discrete and continuous random variables.
We have considered two approaches to construct the causal graphs. 
First, causal graphs are constructed by applying only the domain knowledge of CPS, in particular the knowledge about both the control logic and the physical coupling among the CPS design parameters, like sensors and actuators. 
We have validated these causal graphs using parameter estimation of each of the edges in the graph using historian data logs.
Second, the graphs are learned using historian data logs of a CPS, and no domain knowledge is used in these cases. 
Our analysis shows that the learned graphs have significant similarity with that of domain knowledge based causal graphs. 
Finally, by considering a set of nine well known attacks on SWaT, we have identified the impact points for each of these attacks using causal inference in both the domain knowledge causal graphs and the graphs learned using the historian data logs. 
We have consulted with a plant engineer to validate the impact points of each these attacks.
It turns out that the impact points were either ignored or not known previously, but can be expected in majority of the attacks. 
We believe that our work in this paper  can stimulate a new thread of research works on systematically identifying attack target(s) and attack impact(s) given a set of attack points in a CPS.



\bibliographystyle{ACM-Reference-Format}
\bibliography{biblio}

\appendix





\end{document}